%% file: main.tex
\documentclass[sigplan,10pt]{acmart}

\renewcommand\footnotetextcopyrightpermission[1]{}
\setcopyright{none}
\acmYear{2026}

\makeatletter
\AddToHook{begindocument/end}{%
  \fancypagestyle{standardpagestyle}{%
    \fancyhf{}%
    \renewcommand{\headrulewidth}{\z@}%
    \renewcommand{\footrulewidth}{\z@}%
    \fancyfoot[C]{\if@ACM@printfolios\footnotesize\thepage\fi}%
  }%
  \pagestyle{standardpagestyle}%
}
\makeatother

\usepackage{listings}
\usepackage{tikz}
\usepackage{xspace}
\usepackage{tabularx}
\usepackage{subcaption}
\usepackage{array}
\usepackage{multirow}
\usepackage{multicol}
\usepackage{enumitem}
\usepackage[most]{tcolorbox}

\definecolor{keycolor}{RGB}{0,120,130}
\definecolor{subkeycolor}{RGB}{190,120,60}
\definecolor{strcolor}{RGB}{0,130,20}
\definecolor{commentcolor}{RGB}{110,110,110}
\newcommand{\key}[1]{\textcolor{keycolor}{#1}}
\newcommand{\subkey}[1]{\textcolor{subkeycolor}{#1}}
\newcommand{\str}[1]{\textcolor{strcolor}{#1}}
\newcommand{\com}[1]{\textcolor{commentcolor}{// #1}}
\newtcolorbox{configbox}[1][]{
  enhanced, colback=white, colframe=white, boxrule=0pt,
  left=0pt, right=0pt, top=0pt, bottom=0pt,
  fontupper=\ttfamily\small, #1
}

\input{commenting}
\input{macros}

\sloppypar
\begin{document}


\title{Private Computation Space: Experience with Trusted Multi-Cluster Federated Learning for Agriculture}

\input{authors}

\begin{abstract}
  Artificial Intelligence has shown to help improve agricultural practices, yet adoption remains limited: 69\% of U.S. farmers have privacy concerns with sharing their data, and these concerns must be addressed before adoption is widespread.
  While Federated Learning has been demonstrated to protect privacy at scale for other sectors, deploying a system for agriculture comes with its own set of challenges; the problem necessitates a system that can protect farmer data and identities while preserving model utility, runs on commodity hardware, and is resilient to fragile rural infrastructure.
  To address these concerns, we introduce the \pcsfull, a deployed, open-source Machine Learning system to provision and process farmer data securely. We design a system tailored to an agricultural setting, with multi-cluster orchestration for reliability in rural areas with asynchronous Federated Learning (FL), Differential Privacy (DP), and Trusted Execution Environments (TEEs), to allow farms to participate in the framework while keeping their data private.
  We evaluate the system on two deployed workloads: monitoring nitrogen with living plant sensors in NY for six months and predicting evapotranspiration from weather stations in CA for ten months. Our evaluation finds a \DiceSimilarityCoefficient (\DSC) of \tomatoCDP and $R^2$ accuracy of \etCDP for the respective workloads, improving the worst single-site model accuracy by 22.4\% and 9.1\%, respectively, while preserving privacy.

\end{abstract}

\maketitle


\input{introduction}


\input{background}

\input{experience}

\input{system_design}


\input{implementation}
\input{evaluation}

\input{discussion}

\input{related-work}

\input{conclusion}
\section*{Availability}
The software system artifact, the \pcslong, will be available open-source after the review process. The data used in the system, however, is confidential and will not be released with the artifact.

\bibliographystyle{plain}
\bibliography{ref}

\end{document}

%% file: commenting.tex
\definecolor{WowColor}{rgb}{.75,0,.75}
\definecolor{SubtleColor}{rgb}{0,0,.50}

\ifdefined\Comment
        \renewcommand{\Comment}[1]{}
\else
        \newcommand{\Comment}[1]{}
\fi

\newcounter{margincounter}

\definecolor{ghgreen}{rgb}{0.90,1,0.93}
\definecolor{ghred}{rgb}{1,0.88,0.94}

\definecolor{codegreen}{rgb}{0,0.6,0}
\definecolor{codegray}{rgb}{0.5,0.5,0.5}
\definecolor{codepurple}{rgb}{0.58,0,0.82}
\definecolor{backcolour}{rgb}{0.95,0.95,0.92}

%% file: macros.tex
\usepackage[normalem]{ulem} 

\newcommand{\pcs}[0]{PCS\xspace}
\newcommand{\pcslong}[0]{Private Computation Space\xspace}
\newcommand{\pcsfull}[0]{Private Computation Space (PCS)\xspace}

\newcommand{\sect}[1]{\S\ref{#1}}

\newcommand{\tomatoFL}[0]{0.85\xspace}
\newcommand{\tomatoCentralized}[0]{0.86\xspace}

\newcommand{\tomatoCDP}[0]{0.71\xspace}
\newcommand{\tomatoCDPEPS}[0]{10\xspace}

\newcommand{\DiceSimilarityCoefficient}[0]{Dice Similarity Coefficient\xspace}
\newcommand{\DSC}[0]{DSC\xspace}

\newcommand{\etFL}[0]{0.88\xspace}
\newcommand{\etCentralized}[0]{0.89\xspace}

\newcommand{\etCDP}[0]{0.84\xspace}
\newcommand{\etCDPEPS}[0]{10.49\xspace}

\newcommand{\TEEOverheadBound}[0]{2\%\xspace}

\newcommand{\ETDeployment}[0]{ET deployment\xspace}
\newcommand{\TomatoDeployment}[0]{Nitrogen deployment\xspace}

\newcommand{\ETdeployedTenure}[0]{ten months\xspace}
\newcommand{\TomatodeployedTenure}[0]{six months\xspace}

\newcommand{\annualFoodIncrease}[0]{2.2\%\xspace}

\newcommand{\shuangyu}[1]{\textcolor{magenta}{#1}}

\newcommand{\deleteShuangyu}[1]{}

%% file: authors.tex
%
%

\settopmatter{authorsperrow=5}

\author{Shuangyu Lei}
\affiliation{\institution{Cornell University}\country{USA}}

\author{Muhammad Salman Abid}
\affiliation{\institution{Cornell University}\country{USA}}

\author{Jacob Belding}
\affiliation{\institution{Cornell University}\country{USA}}

\author{Sam Mosher}
\affiliation{\institution{Cornell University}\country{USA}}

\author{Manushi B. Trivedi}
\affiliation{\institution{Cornell University}\country{USA}}

\author{Shivranjani Baruah}
\affiliation{\institution{Cornell University}\country{USA}}

\author{Liam Wickes-Do}
\affiliation{\institution{Cornell University}\country{USA}}

\author{Andrew Anderson}
\affiliation{\institution{IBM Research}\country{USA}}

\author{Braulio Dumba}
\affiliation{\institution{IBM Research}\country{USA}}

\author{Alyssa Whitcraft}
\affiliation{\institution{University of Maryland}\country{USA}}

\author{Ritvik Sahajpal}
\affiliation{\institution{University of Maryland}\country{USA}}

\author{Sijin Li}
\affiliation{\institution{Cornell University}\country{USA}}

\author{Kelly Robbins}
\affiliation{\institution{Cornell University}\country{USA}}

\author{Michael Gore}
\affiliation{\institution{Cornell University}\country{USA}}

\author{Margaret Frank}
\affiliation{\institution{Cornell University}\country{USA}}

\author{Steven Wolf}
\affiliation{\institution{Cornell University}\country{USA}}

\author{Liz Jones}
\affiliation{\institution{Cornell University}\country{USA}}

\author{Abraham Stroock}
\affiliation{\institution{Cornell University}\country{USA}}

\author{Kaitlin Gold}
\affiliation{\institution{Cornell University}\country{USA}}

\author{Hakim Weatherspoon}
\affiliation{\institution{Cornell University}\country{USA}}

%% file: introduction.tex
\section{Introduction}

\newcommand{\growerfootnote}[0]{\footnote{{\em Farmer} is a broad term for someone working the land to produce food, fiber, or fuel, while {\em grower} typically emphasizes cultivating plants and is considered a bit more modern. They are used interchangeably in the paper to refer to people practicing agriculture.}}

\newcommand{\researchfarmfootnote}[0]{\footnote{A {\em research farm} is a university-operated agricultural site where crops, livestock, and farming practices are studied under real field conditions, typically affiliated with a land-grant university~\cite{APLUlandgrant}.}}


As a land-grant research institution, we work directly with growers\growerfootnote.
This paper reports on the experience of two farm- and cloud-level system deployments,
one monitoring nitrogen levels via sentinel plants in New York and
a second predicting evapotranspiration (ET) using weather station data from the California Irrigation Management Information System (CIMIS) \cite{cimis_stations2025}.
These two deployments highlight increasingly common Machine Learning (ML) workloads where farmers contribute data to help manage on-farm resources.
However, a key fundamental issue prevents farmers from participating: privacy. Slattery et al. report that in the U.S., 69\% of farmers had privacy concerns when asked to share their data, 73\% do not trust private companies, and 58\% do not trust the government \cite{trustinfood2021farmerdata}.
This necessitates a system that growers can \emph{trust} with their data and make use of it to apply Digital Agriculture to their operation with privacy guarantees.

\Comment{
    \shuangyu {Do we need to repeat the two workloads' descriptions?}
    In this spirit, we design the \pcsfull as a federated system and operate it on two agricultural workloads. Together, the two deployments yield the operational lessons we report in \sect{sec::lessons}.
}

\Comment{
    The United Nations estimates that the global population will grow to nearly 10 billion people by 2050 \cite{FAO2009}. With such growth, there needs to be a reliable, systemic way to feed the global population that balances the needs of smallholder farms with large agrifood producers\Comment{, and recent studies show that only a \annualFoodIncrease increase in food production per year\footnote{A 70\% increase in production by 2050 is an annual increase by 2.2\% of 25 years.} can get us \Comment{closer }to that goal} \cite{FAO2009, unfss2_highlights}.
    The use of \Comment{Machine Learning (ML) and }Artificial Intelligence (AI) plays an essential role in scientific computation to model and predict resource allocation to improve crop efficiency and yield \cite{iqbal2025sustainable}.
    With the use of AI, however, there is a strong requirement to guarantee privacy for the farmers\growerfootnote and service providers contributing data to these analytical programs. Slattery et al. report that in the U.S, 69\% of farmers had privacy concerns when asked to share their data, 73\% do not trust private companies, and 58\% do not trust the government \cite{trustinfood2021farmerdata}. This necessitates a system that growers can \emph{trust} with their data and make use of it to apply Digital Agriculture to their operation with privacy guarantees.
}

We introduce the \pcsfull, a deployed system that runs agricultural workloads farmers can \textit{trust}, accommodates poor connectivity in rural settings, and generalizes across workloads. \pcs consists of two parts: (i) a Kubernetes-based multi-cluster orchestration framework for running workloads on on-premise farm clusters and the public cloud,
and (ii) an asynchronous Federated Learning framework for training ML models on sensitive farmer data with Differential Privacy.
With a combination of the two, farmer data can be reliably provisioned and processed in a hybrid setting;
farm clusters run on-premises so data does not leave the client premises during training, and model weights are aggregated in the public cloud within a Trusted Execution Environment (TEE).

As a distributed ML approach, Federated Learning (FL) enables training on a large corpus of decentralized data residing on devices \cite{bonawitz2019googlefl}.
FL has been deployed across agricultural applications to solve privacy concerns \cite{dembani2025agricultural}.
To further secure FL systems, prior works have used three privacy-preservation schemes: Fully-Homomorphic Encryption, Secure Multi-Party Computation, and Differential Privacy \cite{dembani2025agricultural}.
While individual privacy-preserving FL practices have made considerable progress, existing deployments lack general support for heterogeneous applications.
Selecting an optimal scheme for a general agricultural framework requires considering its complex threat model, data and application heterogeneity, limited on-farm resources, and intermittent network partitions during the FL system's lifecycle (\sect{sec::experience}).

The FL framework in \pcs is inspired in design by other large-scale FL frameworks, including Papaya FL from Meta \cite{huba2022papaya}, Google FL \cite{bonawitz2019googlefl} and Apple FL \cite{paulik2021applefl}.
Although terminology is reused from the more recent Papaya FL system for consistency, \pcs implements both SyncFL and AsyncFL (FedBuff \cite{nguyen2022federatedlearningbufferedasynchronous}) and is
evaluated extensively in agricultural settings.
To strengthen the privacy guarantees, \pcs also applies \emph{Differential Privacy} to anonymize user participation in the FL framework and \emph{Trusted Execution Environments} to secure aggregation from unverified aggregator and infrastructure providers. In comparison, Papaya FL relies on random masking, which does not protect participants from membership inference attacks \cite{shokri2017membership}.
While existing large-scale FL frameworks target mobile and consumer settings, FL has to be specialized for domains with stricter privacy requirements and deployment setups. Systems such as Fed-BioMed \cite{Cremonesi2025} work towards a specialized setup for healthcare, whereas \pcs targets farms and has been deployed on several research farms.
A {\em research farm} is an agricultural site where crops, livestock, and farming practices are studied under field conditions, typically affiliated with a land-grant university~\cite{APLUlandgrant}.
This paper discusses the implications of designing for agriculture (\sect{sec::experience}) and operational metrics from current long-running deployments (\sect{sec::evaluation}).


We deploy \pcs in two long-running agricultural settings: (i) monitoring nitrogen levels across three farms in New York for six months, and (ii) predicting
evapotranspiration across ten weather stations in California for ten months.
Across both deployments, FL improves over single-site training by leveraging geographically distributed and heterogeneous datasets while keeping raw data local. When applying Differential Privacy (DP) at the aggregator, called Central DP, \pcs retains strong model utility. It achieves a measure of accuracy called Dice Similarity Coefficient (DSC) of 0.71, which improves the accuracy of the worst single-site model by 22.4\% for nitrogen monitoring, and an $R^2$ of 0.84 for ET prediction, improving the accuracy of the worst single-station model by 9.1\%.
Over deployment periods of six and ten months, respectively, \pcs continues operating client clusters on Raspberry Pis successfully despite client dropouts and network disruptions. Compared to a conventional Kubernetes deployment, TEE-secured aggregation introduces less than \TEEOverheadBound overhead relative to end-to-end training time. These results demonstrate that privacy-preserving FL can be deployed in agricultural environments without sacrificing privacy, practicality, reliability, or model quality.

To summarize our key contributions, we:
\begin{itemize}[itemsep=1pt, topsep=2pt, parsep=0pt, partopsep=0pt]
    \item Deploy a trusted multi-cluster FL framework for agriculture that ensures privacy by design. 
    \item Report our experience, operational realities, and lessons learned with a deployed system that protects growers' privacy, with flexibility for rural internet connections.
    \item Describe deployment results, reporting significant accuracy improvements compared with single-site training and low latency overhead for both workloads.
\end{itemize}



%% file: background.tex
\section{Background}
\label{sec::background}

Digital Agriculture (DA) is the practice of applying digital methods and technologies to traditional agricultural practices \cite{tilman2011global, Mueller2012, vasisht2017farmbeats}.
A prominent example is precision agriculture, which collects data specific to a region, or precise location, to improve resource management through precise application of inputs such as water, fertilizer, and feed \cite{gold2021plant,space2farmNews}.
The U.S. Department of Agriculture reports that 68\% of large-scale crop-producing farms used yield monitors, yield maps, and soil maps in 2023 \cite{precisionAgERS}.


Critical functions of farmer-facing DA tools are:
\begin{itemize}[itemsep=1pt, topsep=2pt, parsep=0pt, partopsep=0pt]
    \item \textit{Use fewer resources to grow crops}: effectively irrigating and fertilizing fields with less, benefiting the environment and improving long-term profitability \cite{precisionAgGAO}.
    \item \textit{Predict disease before it happens}: supporting decisions that curb widespread disease across farms \cite{MLMethods, MLpipeline}.
    \item \textit{Operate reliably on the farm}: rural regions have unreliable internet connectivity, and any solution must tolerate network partitions \cite{comosum}.
    \item \textit{Keep grower data private}: sensitive information about crop yield or disease could be exploited by competitors or third parties, harming the grower's business \cite{wilgenbusch2022addressing}.
\end{itemize}

\subsection{Machine Learning for DA}
Modern agricultural hardware monitors farms through precise irrigation \cite{Almarshadi2011EffectsOP}, soil analysis \cite{Kim2009SoilMS}, and plant-health sensors. Machine Learning extends this monitoring with prediction: models trained on the collected data can forecast crop conditions ahead of time, supporting proactive maintenance and decisions for better crop health \cite{MLpipeline}. Research on agricultural ML is ongoing, and we discuss it in more detail in \sect{sect::related}.
\begin{figure}[t]
    \begin{center}
        \includegraphics[width=\linewidth]{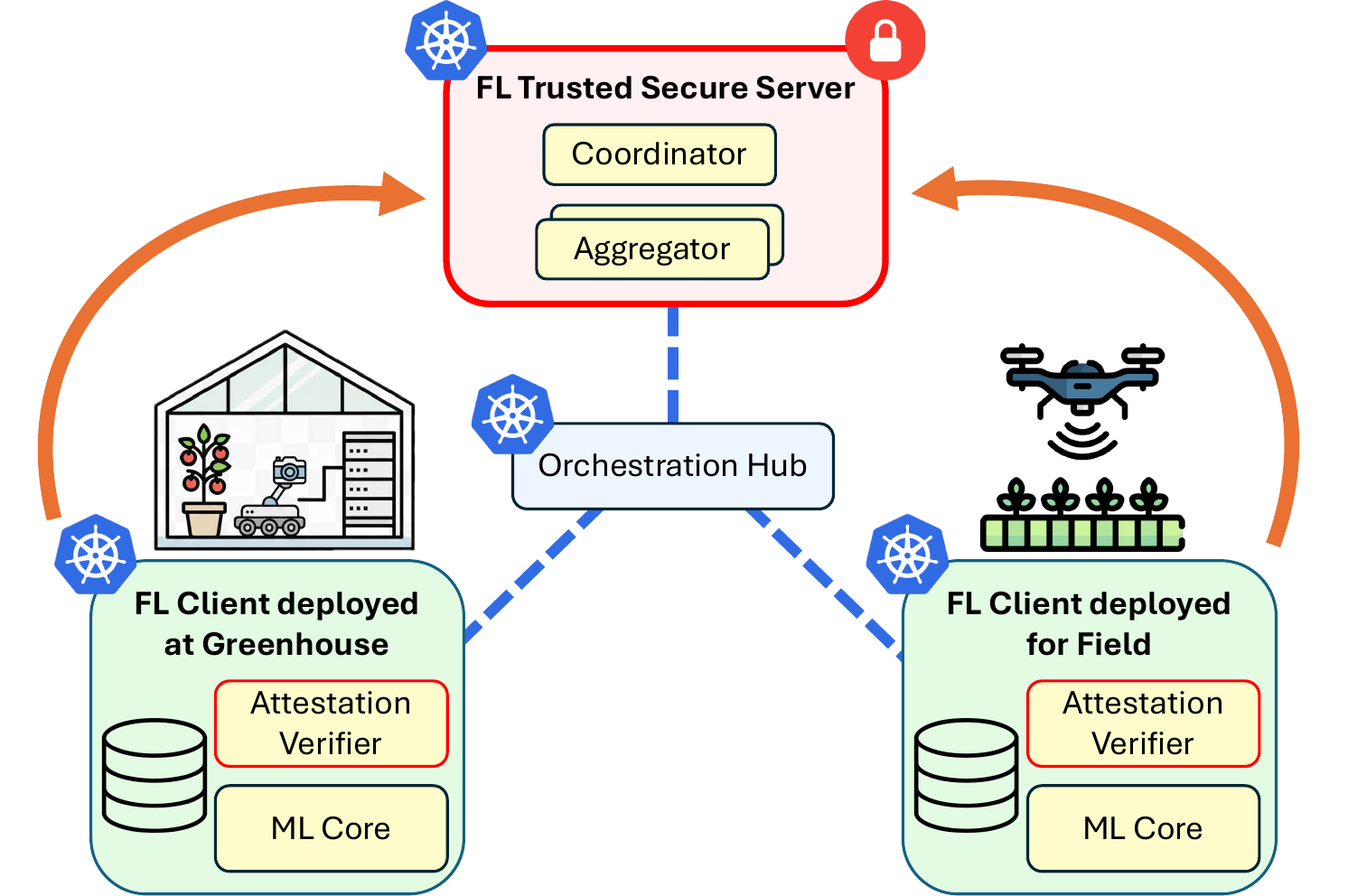}
    \end{center}
    \caption{\label{fig:tomato-deploy} Multi-cluster Federated Learning deployment for nitrogen estimation with sentinel plants in New York.}
\end{figure}

\begin{figure}[t]
    \begin{center}
        \includegraphics[width=\linewidth]{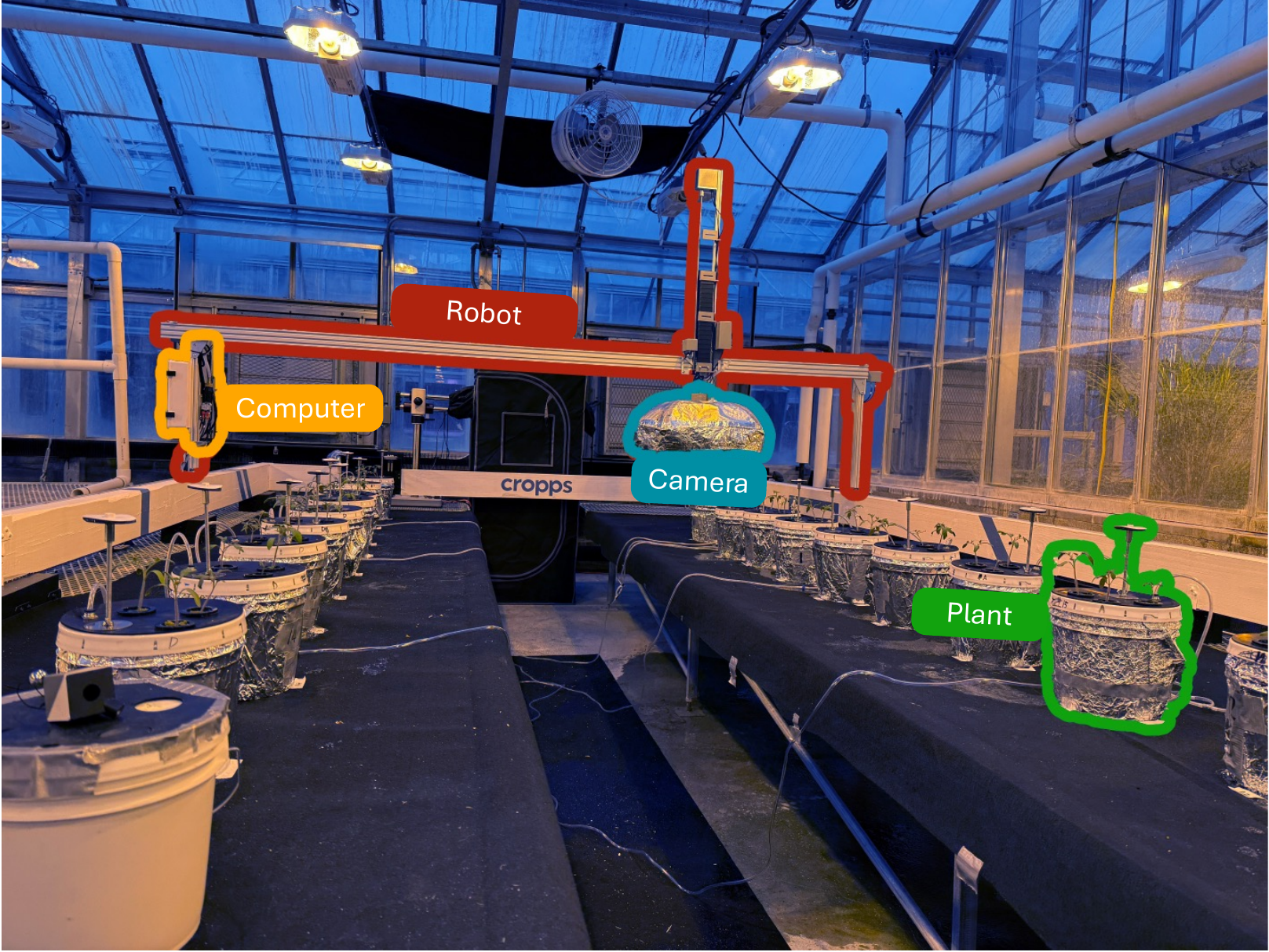}
    \end{center}
    \caption{\label{fig:tomato-farmbot} FarmBot \cite{farmbot} imaging setup at a greenhouse client.
        The robot carries a camera that captures each of the 54 tomato plants in the bench.}
\end{figure}

\begin{figure}[th]
    \begin{center}
        \includegraphics[width=\linewidth]{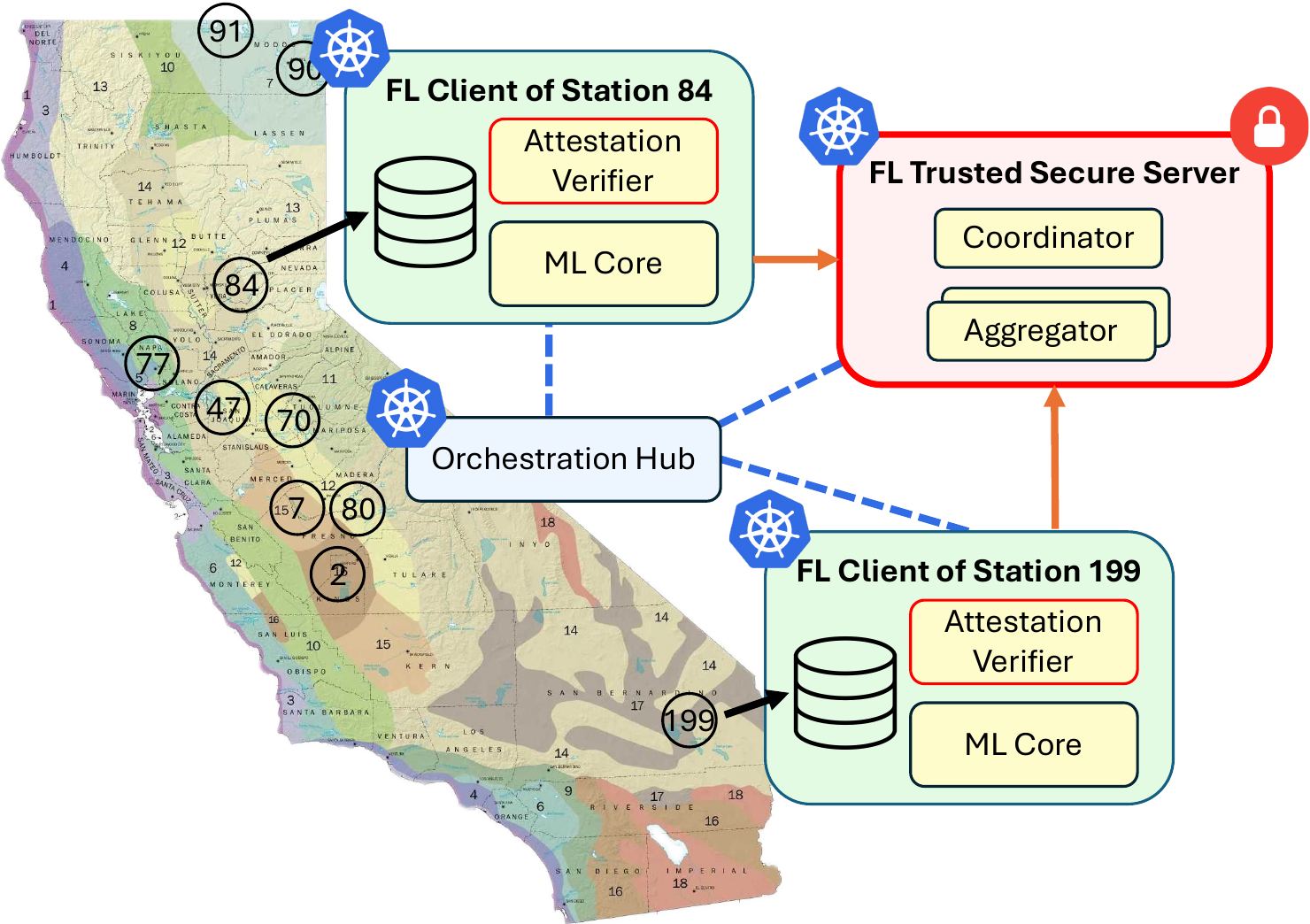}
    \end{center}
    \caption{\label{fig:et-deploy} Multi-cluster Federated Learning deployment for ET prediction across California. Ten climatically distinct weather stations (labeled by station ID) run as FL clients. Map colors correspond to ET zones \cite{cimis_stations2025}.}
\end{figure}
\subsection{Privacy Concerns in DA}
A prominent challenge in Digital Agriculture research is maintaining user privacy while working with confidential data \cite{trustinfood2021farmerdata}.
This challenge is an obstacle to the widespread adoption of ML models, especially for both smallholder and large-scale farmers concerned about privacy.
For farmers, the privacy concern centers on identifying information in their sensor data, such as coordinates, location metadata, and embedded owner information.
Farmers have three principal reasons to keep this information private.
First, \textit{the right to privacy}: farmers may not consent for their information to be used to identify them or their farm by any assistive software, and current literature documents instances of data misuse by ML or AI models \cite{altman2024techbrief}.
Second, \textit{financial risk}: data leaks may lead to financial repercussions or allow other companies to profit from this information \cite{wilgenbusch2022addressing, trustinfood2021farmerdata}.
Third, \textit{equitable participation}: privacy control allows smallholder farmers to engage in food production on more equal footing with larger producers by retaining the value of their own data \cite{wolf1997privatization}.

\subsection{Privacy with Secure Hardware}

To keep aggregation data secure, some existing FL systems use Multi-Party Computation \cite{bonawitz2019googlefl, so2021turbo} or Trusted Execution Environments \cite{huba2022papaya} with a secondary privacy measure, e.g., masking individual client reports in Papaya FL. Considering the computational requirements for using multi-party computation, we select TEEs for securing user data on the cloud. In a TEE, the initial state of the machine and the software running on the node can be measured with \textit{remote attestation}; remote attestation measures the state of the VM, kernel, boot configuration, and deployed binary, and generates an \textit{Attestation Quote} (AQ). This quote is cryptographically signed by the TEE, providing strong guarantees that it cannot be forged or altered by malicious actors.
The generated AQ can then be compared with a reference hash of the open-source binary running on the remote server, establishing the TEE as a trusted platform to process confidential data. We use Confidential VMs (CVMs) to provision TEEs on demand, and discuss our application of CVMs for FL in \sect{sec::implementation:tee}.




\subsection{Synchronous and Asynchronous FL}

\begin{figure*}[th]
    \begin{center}
        \includegraphics[width=\textwidth]{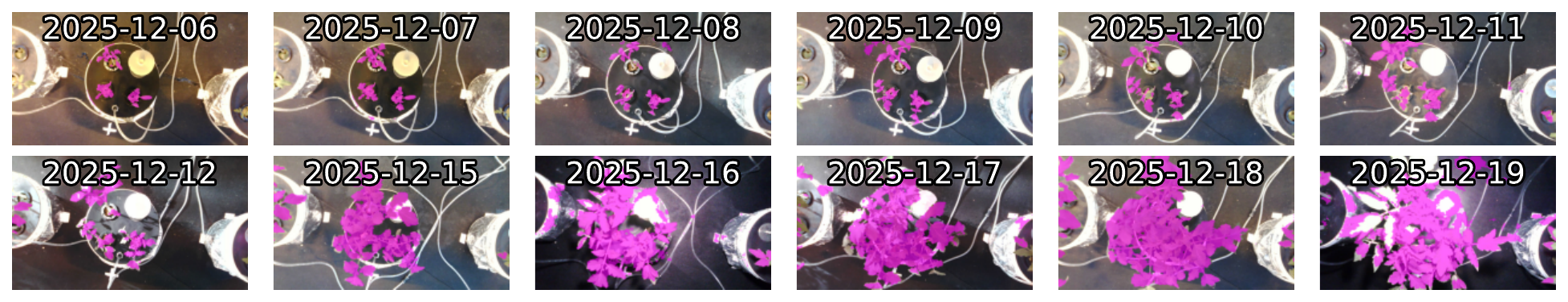}
    \end{center}
    \caption{\label{fig:tomato-deploy-log} \textbf{\TomatoDeployment}: Federated segmentation of one greenhouse plant over a 14-day cycle.}
\end{figure*}

\begin{figure}[th]
    \begin{center}
        \includegraphics[width=\linewidth]{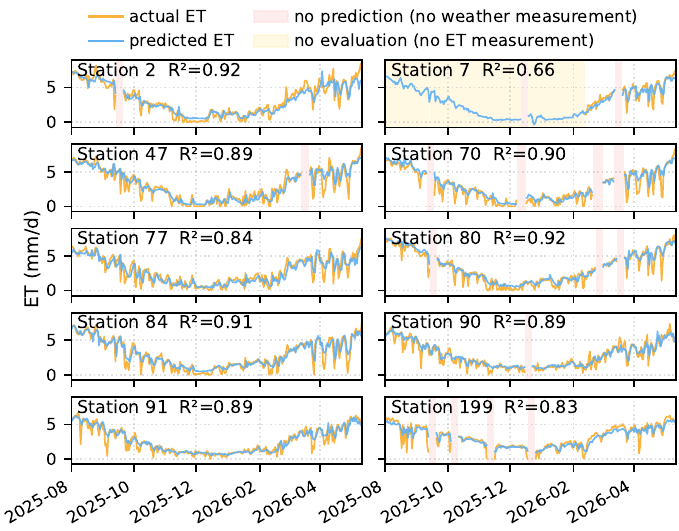}
    \end{center}
    \caption{\label{fig:et-deploy-log} Daily live ET predictions versus station-measured ET across the ten CIMIS stations for \ETdeployedTenure.}
\end{figure}

Federated Learning is a decentralized training method. Each client trains locally and sends only model updates to a central server, which aggregates them into a refined global model.
In synchronous FL (SyncFL), the server waits for updates from all participating clients before aggregating \cite{fedavg}.
SyncFL is straightforward to analyze but is prone to the straggler problem: each round proceeds at the pace of the slowest client, so dropouts or partitions stall progress.
Asynchronous FL (AsyncFL) avoids this problem by aggregating updates as they arrive rather than waiting for all clients.
FedAsync \cite{chen2020asynchronous} updates the global model on every client upload, but the per-update aggregation incurs heavy server cost.
FedBuff \cite{nguyen2022federatedlearningbufferedasynchronous} mitigates this by aggregating $K$ client updates per server step, reducing aggregation frequency while preserving AsyncFL's tolerance of stragglers and partitions.

\subsection{Differential Privacy}
Differential Privacy is a mathematical way to bound how much an attacker can learn about any single record from a computation's output \cite{dp}.
DP adds calibrated noise to hide any single record's contribution.
Privacy budget $\epsilon$ quantifies the privacy guarantee of DP: for any two datasets differing in one record, the probability of producing the same output differs by at most a factor controlled by $\epsilon$. Smaller $\epsilon$ means stronger privacy.

%% file: experience.tex
\section{Experience}
\label{sec::experience}

\begin{table*}[th]
    \centering
    \setlength{\tabcolsep}{4pt}
    \renewcommand{\arraystretch}{1.2}
    \begin{tabular}{|p{0.06\textwidth}|p{0.28\textwidth}|p{0.27\textwidth}|p{0.25\textwidth}|}
        \hline
        \textbf{Reality} & Complex threat model and privacy goals                                                         & Heterogeneous workloads and constrained privacy mechanisms          & Fragile farm infrastructure and scarce heterogeneous data                      \\
        \hline
        \textbf{Lesson}  & Privacy preservation requires combining FL, TEE-secured aggregation, and Differential Privacy. & Generalizable framework requires model-agnostic privacy techniques. & Resilient deployment requires multi-cluster orchestration and asynchronous FL. \\
        \hline
    \end{tabular}
    \caption{Three operational realities and lessons learned of privacy-preserving agricultural deployments.}
    \label{tab:realities-lessons}
\end{table*}


As a land-grant research institution, we work directly with growers.
In collaboration with the National Aeronautics and Space Administration (NASA) and a National Science Foundation (NSF) research center,
we design \pcs and operate two long-running system deployments in agriculture:

\begin{itemize}[itemsep=1pt, topsep=2pt, parsep=0pt, partopsep=0pt]
    \item \textbf{Estimating nitrogen via sentinel plants in NY} (\sect{sec::ny-deployment}). \pcs runs a federated image analysis pipeline across greenhouses and field site of an NSF research center to monitor soil nitrogen levels through sentinel plants.
          The deployment has run for \TomatodeployedTenure.
    \item \textbf{Predicting evapotranspiration in CA}  (\sect{sec::ca-deployment}). \pcs predicts next-day ET via a federated model across ten California weather stations spanning multiple climate zones. The deployment has run for \ETdeployedTenure.
\end{itemize}

\subsection{Deployment 1: Soil nitrogen monitor}
\label{sec::ny-deployment}

Nitrogen monitoring is essential for fertilizer management but traditionally requires lab-based soil tests.
The NSF research center has developed genetically engineered sentinel plants that visually indicate nitrogen deficiency through leaf color.
These plants are distributed across research farms in New York as spatially distributed soil-nitrogen sensors.
For \TomatodeployedTenure, \pcs has run an automated four-stage image analysis pipeline: plant segmentation, color reference segmentation, color correction, and nitrogen estimation. The plant segmentation model is trained through Federated Learning.

The deployment has FL clients across the greenhouses and field site, each processing its local image dataset and training the plant segmentation model on-site (Figure \ref{fig:tomato-deploy}).
Greenhouse imaging is automated by a robot that captures plant images on a fixed schedule or upon request (Figure~\ref{fig:tomato-farmbot}).

The six-month deployment spans multiple sentinel-plant experiment cycles.
Figure~\ref{fig:tomato-deploy-log} shows the deployed federated model successfully segmenting one greenhouse plant across all growth stages over a 14-day cycle.
In \sect{sec::evaluation}, the federated model keeps raw images on-site and achieves \DSC$=\tomatoFL$, indicating close agreement between predicted and ground-truth masks.
Applying Central DP at $\epsilon=\tomatoCDPEPS$ reduces \DSC to $\tomatoCDP$ in exchange for a formal membership privacy guarantee.
DP limits how much an attacker can learn about whether any specific image was part of training, even with full access to the trained model.

\subsection{Deployment 2: Evapotranspiration prediction}
\label{sec::ca-deployment}

Effective irrigation scheduling requires accurate evapotranspiration predictions. ET is the combined loss of water through soil evaporation and plant transpiration.
The California Department of Water Resources operates the California Irrigation Management Information System \cite{cimis_stations2025}, a public network of over 270 automated weather stations supporting irrigation decisions for water-stressed farmlands across the state.
\pcs deploys ten FL clients across distinct climate zones, each representing a different CIMIS station.
For \ETdeployedTenure, \pcs has trained a federated model that predicts next-day ET from a window of recent weather measurements.





The deployment has ten FL clients, each pulling data from a CIMIS station via the public API and producing next-day ET predictions continuously (Figure \ref{fig:et-deploy}).
The federated model was trained on 12 years of historical CIMIS data.


In ten months of live prediction, the federated model achieves per-station $R^2$ between 0.84 and 0.92 for 9 of the 10 stations (Figure~\ref{fig:et-deploy-log}), explaining most of the variance in measured ET; the one outlier ($R^2 = 0.66$) had no ground truth ET measurements for 7 months of the deployment.
In \sect{sec::evaluation}, the federated model achieves $R^2=\etFL$ on a 5-year test set while keeping raw data on-site.
Applying Central Differential Privacy at $\epsilon=\etCDPEPS$ reduces $R^2$ to $\etCDP$ in exchange for a formal guarantee of membership privacy.

\subsection{Operational Realities}
\label{sec::challenges}

Across two deployments developed in direct collaboration with farmers, the major questions farmers care about are:

\begin{itemize}[itemsep=1pt, topsep=2pt, parsep=0pt, partopsep=0pt]
    \item \textbf{Accuracy}: How effectively can ML models assist in making actionable decisions?
    \item \textbf{Privacy}: Can the framework protect privacy without sacrificing model accuracy?
    \item \textbf{Reliability}: Can the system work reliably with fragile infrastructure at the farm?
    \item \textbf{Overhead}: Does adding privacy-preserving techniques in the system require more resources?
\end{itemize}

Addressing these concerns in an agricultural deployment surfaces three operational realities that shape \pcs' design.

\subsubsection{Complex threat model and privacy goals}
\label{sec:threat_model_and_goals}

The data involved in \pcs is private by nature. For the \TomatoDeployment, plant imagery reveals nitrogen stress, disease, and treatment conditions \cite{MLpipeline}.
Although the \ETDeployment uses public CIMIS data, the same principle would apply to farmer-owned weather sensors as part of the system;
the weather record can reveal identifying irrigation schedules and crop management practices.
Both types of information are sensitive as they can draw attention from competitors or affect crop pricing \cite{wilgenbusch2022addressing, 10.3389/fsufs.2022.903230}.

To protect farmers' privacy, \pcs adopts an honest-but-curious threat model for all three stakeholders: farmers, model developers, and infrastructure providers.
This threat model follows the convention in privacy-preserving analytics and learning \cite{mycelium, arboretum, huba2022papaya, bonawitz2019googlefl}, and fits agricultural settings. We assume participants are not malicious because they use the model themselves and poisoning training would degrade their own predictions.
However, participants may be curious and attempt to infer farmers' sensitive data or contributions.
To defend against these inferences, the system must satisfy three privacy properties:
\begin{itemize}[itemsep=1pt, topsep=2pt, parsep=0pt, partopsep=0pt]
    \item \textbf{Data privacy}: training must not reveal a client's data to other clients or third parties.
    \item \textbf{Computation privacy}: the aggregation must not reveal which clients participated or what they contributed.
    \item \textbf{Membership privacy}: the released model must not leak if any individual record was in the training set.
\end{itemize}

Side-channel and hardware attacks on the TEE and physical compromise of farm devices are out of scope.

\subsubsection{Heterogeneous workloads and constrained privacy mechanisms}
\label{sec::workload_heterogeneity}

Agricultural workloads vary widely in model architecture and data structure, including image segmentation for the \TomatoDeployment, time-series regression for the \ETDeployment, and future tasks.
\pcs must provide practical and generalizable solutions across heterogeneous workloads.

Several privacy mechanisms are considered for \pcs but are ruled out in agricultural settings.
Some are operationally infeasible on the farm: Fully Homomorphic Encryption is too computationally expensive to run complex models on farm devices \cite{dembani2025agricultural}; Secure Multi-Party Computation \cite{bonawitz2019googlefl, so2021turbo} requires multi-round client coordination and is sensitive to dropouts, making it infeasible on rural networks.
Others must be tailored to each workload: applying DP to client data directly requires specialized noising algorithms aware of the data structure and model architecture.
Otherwise, data-level noise destroys structural signals such as spatial correlations in imagery \cite{MLpipeline} and temporal correlations in weather time series.
Instead, applying DP to model weights is more generalizable and retains model accuracy across heterogeneous workloads.

\subsubsection{Fragile farm infrastructure and scarce heterogeneous data}
\label{sec::farm_constraints}
Privacy-preserving ML deployments on farms face two operational constraints: fragile infrastructure and scarce heterogeneous data.
Farm networks are intermittent and bandwidth-limited \cite{dembani2025agricultural, comosum, vasisht2017farmbeats}, and farm devices vary in compute capability.
Together, these cause client dropouts and straggler effects during training.

In addition, farm data is scarce and structurally heterogeneous (non-IID). Data collection is costly and time-intensive, leaving each client with a small dataset that cannot train a robust model on its own. Each client also captures different conditions that other clients may not, so every client's participation contributes meaningfully to shared model robustness. For the \TomatoDeployment, imagery varies in plant growth stage, camera angle, and background across clients; for the \ETDeployment, stations span different climates with varying availability and history depth. As a result, large-scale FL strategies that assume a large client pool and abundant data do not apply. For example, subsampling trains on a small fraction of devices per round, and over-selection sends tasks to extra clients and ignores late responses \cite{huba2022papaya, bonawitz2019googlefl, paulik2021applefl}.

\subsection{Lessons Learned}
\label{sec::lessons}

Deploying a privacy-preserving FL system in two long-running systems has highlighted three significant lessons, summarized in Table~\ref{tab:realities-lessons} and below.

\subsubsection{Privacy preservation requires combining FL, TEE, and DP}
\pcs combines three mechanisms, each addressing a different privacy property and actor. Federated Learning keeps raw data on-device. TEE-secured aggregation isolates the computation via memory encryption, hiding client updates and model weights from the infrastructure provider. DP adds noise to model weights before release, bounding the modeler's ability to infer individual training records from the trained model.


\subsubsection{Generalizable framework requires model-agnostic privacy techniques}
\pcs's privacy stack is practical and generalizable: it neither degrades performance nor requires modifying the model architecture. Federated learning with Central DP, which adds noise to model weights during aggregation, yields good model accuracy, improving significantly over single-site training while protecting privacy. The TEE requires specialized hardware on the cloud but does not constrain client devices or introduce significant overhead to end-to-end training time. Although TEE memory is more costly than commodity memory, the server keeps the TEE memory footprint small by including only the aggregation logic and model weights, without a bundled ML framework. 
Consequently, the platform provides a uniform modeler-side interface: modelers submit federated training requests, and \pcs handles aggregation and exposes training metrics without leaking raw data.

\subsubsection{Resilient deployment requires multi-cluster orchestration and asynchronous FL}
As deployments scale across farm sites, manually deploying and restarting workloads after crashes becomes infeasible. Each farm cluster uses a single containerized binary with outbound-only communication, removing the need for NAT, firewall configuration, or static IPs. Farm clusters tolerate network partitions and resume from their last state on reconnect, without manual redeployment after outages.

To ensure progress despite client dropouts within each training task, \pcs uses FedBuff, which aggregates $K$ client updates per round, with $K$ chosen empirically for each workload. Larger $K$ reduces variance across clients but slows progress under partitions. Under DP, $K$ must lie in a window: large enough to average DP noise but small enough to avoid stalling. Without DP, smaller $K$ does not materially affect final accuracy; it introduces variance in initial rounds but ensures progress under severe partitions.


\begin{figure}[t]
    \begin{center}
        \includegraphics[width=\linewidth]{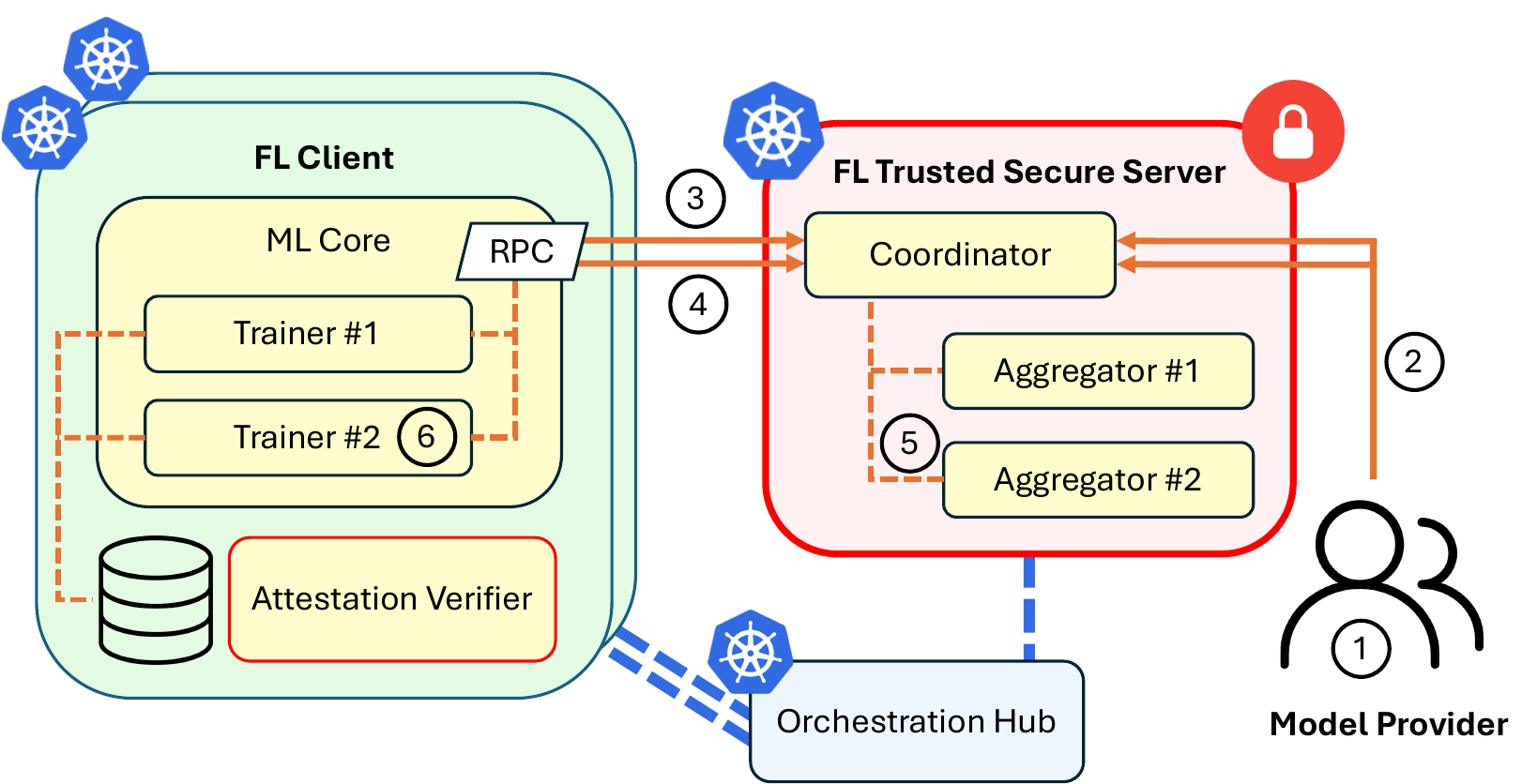}
    \end{center}
    \caption{\label{fig:design} Overall design diagram for \pcs with multi-cluster orchestration.
        Each training task (model, config) spawns an aggregator-trainer pair, each aggregator has a one-to-one relationship with a trainer downstream. }
\end{figure}

Our key insight to design a private, practical, and reliable system is to bring the code to the data via FL, secure server-side computation with TEE, and protect client participation with Differential Privacy. With multi-cluster orchestration and flexibility for rural internet connections, the system can serve as a platform-as-a-service for any agricultural privacy-preserving FL workload.

%% file: system_design.tex

\section{System Design}
\label{sec::design}

\begin{figure}[t]
    \begin{center}
        \includegraphics[width=\linewidth]{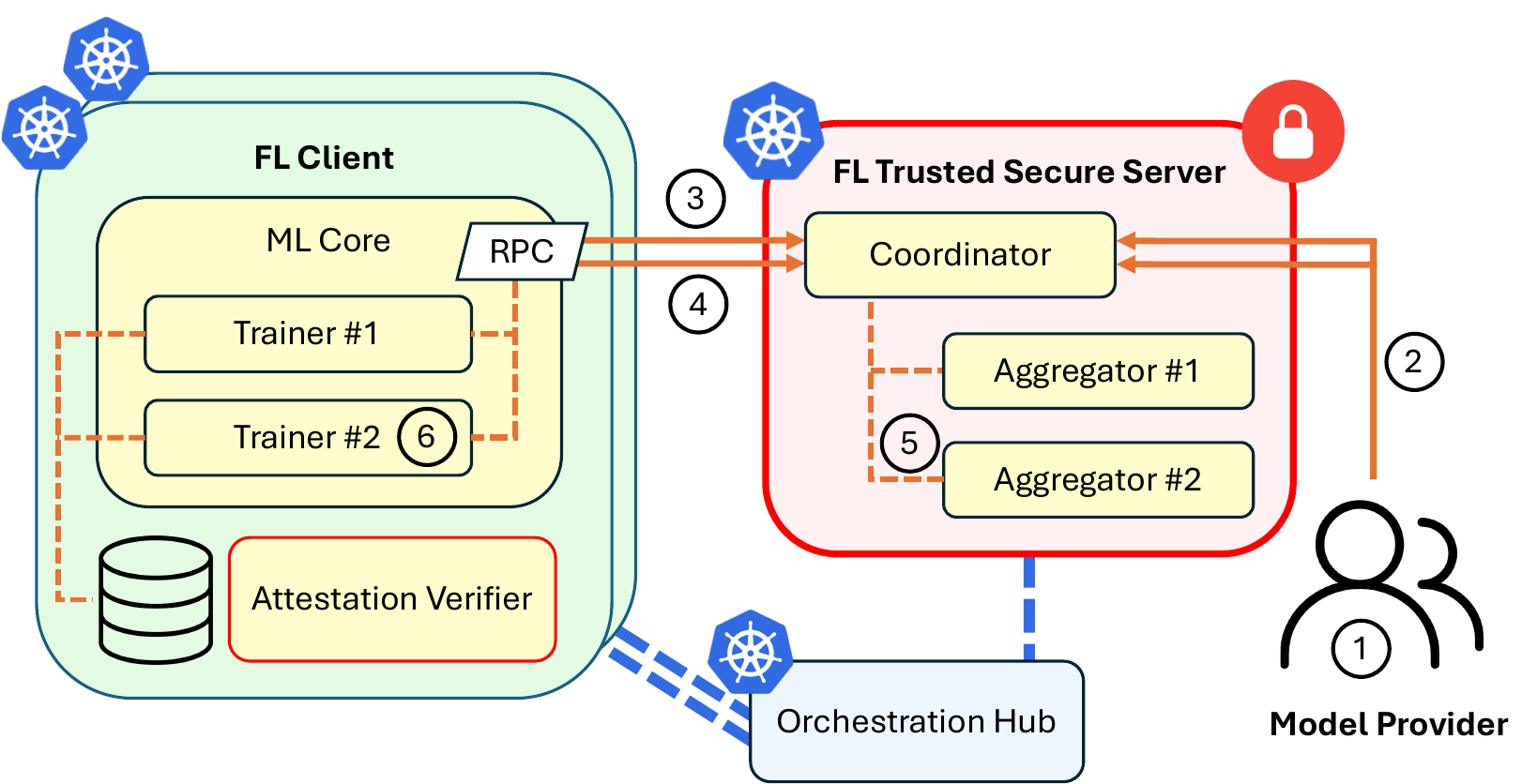}
    \end{center}
    \caption{\label{fig:design} Overall design diagram for \pcs with multi-cluster orchestration.
        Each training task (model, config) spawns an aggregator-trainer pair, each aggregator has a one-to-one relationship with a trainer downstream. }
\end{figure}

\begin{figure}[t]
    \centering
    \begin{configbox}
        \begin{tabular}{@{}l@{\hspace{1em}}l@{}}
            \key{model}:                         &                                   \\
            \quad \subkey{type}: \str{lstm}      &                                   \\
            \quad \subkey{hidden}: 64            &                                   \\

            \key{aggregation}:                   &                                   \\
            \quad \subkey{method}: \str{fedavg}  &                                   \\
            \quad \subkey{mode}: \str{async}     & \com{fallback from sync}          \\
            \quad \subkey{buffer}: 8             & \com{FedBuff $K$ parameter}       \\

            \key{privacy}:                       &                                   \\
            \quad \subkey{mode}: \str{centralDP} & \com{applied at TEE}              \\
            \quad \subkey{epsilon}: 10           & \com{total privacy budget}        \\
            \quad \subkey{delta}: 1e-5           & \com{privacy failure probability} \\
            \quad \subkey{clip}: 0.5             & \com{gradient clipping bound}
        \end{tabular}
    \end{configbox}
    \vspace{-8pt}
    \caption{Example federated request for \ETDeployment.}
    \label{fig:fed-config}
\end{figure}

To build a generalizable privacy-preserving platform for digital agriculture, the system (i) keeps raw data processing on client premises with Federated Learning, (ii) secures off-client computation within a Trusted Execution Environment, and (iii) noises individual contributions to the aggregated model with Differential Privacy. For reliable operation under fragile rural infrastructure \cite{Rubambiza2022seamless}, \pcs uses multi-cluster orchestration to deploy client and server clusters, and support both synchronous and asynchronous FL aggregation to tolerate stragglers and client dropouts.

The system comprises three components: (1) clients on farm clusters, (2) a TEE-based Trusted Secure Server (TSS) on the public cloud or a data center, and (3) an orchestration hub on a cloud server. A modeler derives an aggregated model from distributed datasets through the workflow in Figure \ref{fig:design}:
\begin{enumerate}[itemsep=2pt, topsep=2pt, parsep=0pt, partopsep=0pt]
    \item The modeler authors a federated request as a JSON file (Figure \ref{fig:fed-config}) with two parts: a base model configuration for local training, and an aggregation specification with methods and privacy techniques.
    \item The modeler publishes the request to the TSS coordinator, which dispatches it to clients and monitors progress.
    \item Each client downloads the request. Before training locally, it requests an Attestation Quote (AQ) from the TSS to verify the server, then decides whether to accept based on local availability and privacy rules.
    \item The client uploads model weights or gradients to the TSS over a secure channel.
    \item The TSS aggregates updates across clients following the request specification, then releases the result to clients or the modeler.
    \item The client evaluates the aggregated model on local data and makes predictions on the live data stream.
\end{enumerate}

\subsection{Privacy}

The system protects privacy through three mechanisms: FL, TEE, and DP.




\subsubsection{Protect Data on Client Clusters via FL}

With Federated Learning, each client processes data locally and shares only model weights with the server, so raw data never leaves the farm.
The client consists of:
\begin{itemize}[itemsep=1pt, topsep=2pt, parsep=0pt, partopsep=0pt]
    \item \textbf{Local database} that securely stores data at rest and ingests continuous data from the farm.
    \item \textbf{Machine Learning Core} that runs training and inference. The client pulls federated requests from the server, which specify the model type, data range, and privacy-related parameters. The client can reject requests that violate local policies (e.g., privacy constraints or insufficient data).
    \item \textbf{Attestation Verifier} that requests an Attestation Quote from the Trusted Secure Server and verifies its authentication, detailed in \sect{sec:TEE-based_TSS}.
\end{itemize}

\subsubsection{Secure Computation on Server cluster via TEE}
\label{sec:TEE-based_TSS}

The Trusted Secure Server handles the aggregation logic that combines local model weights from clients. Hosting the server within a Trusted Execution Environment secures aggregation against the model provider, the infrastructure provider, and any third parties.

The TSS consists of the following components:

\textbf{Coordinator} manages FL tasks and monitors clients and servers. For each training task, it spawns an aggregator and selects the FL algorithm (synchronous or asynchronous) based on client availability and task specification. To make use of all client updates to maximize model performance, a synchronous FL aggregation method like FedAvg \cite{fedavg} is used by default. Given unstable farm networks, AsyncFL (FedBuff \cite{nguyen2022federatedlearningbufferedasynchronous}) serves as a fallback to handle dropouts and stragglers. The coordinator also detects tasks stalled by network partitions and reinitiates them once connectivity returns.

\textbf{Aggregator} executes the concrete aggregation logic for a training task following the modeler's FL request.

\begin{figure}
    \begin{center}
        \includegraphics[width=\linewidth]{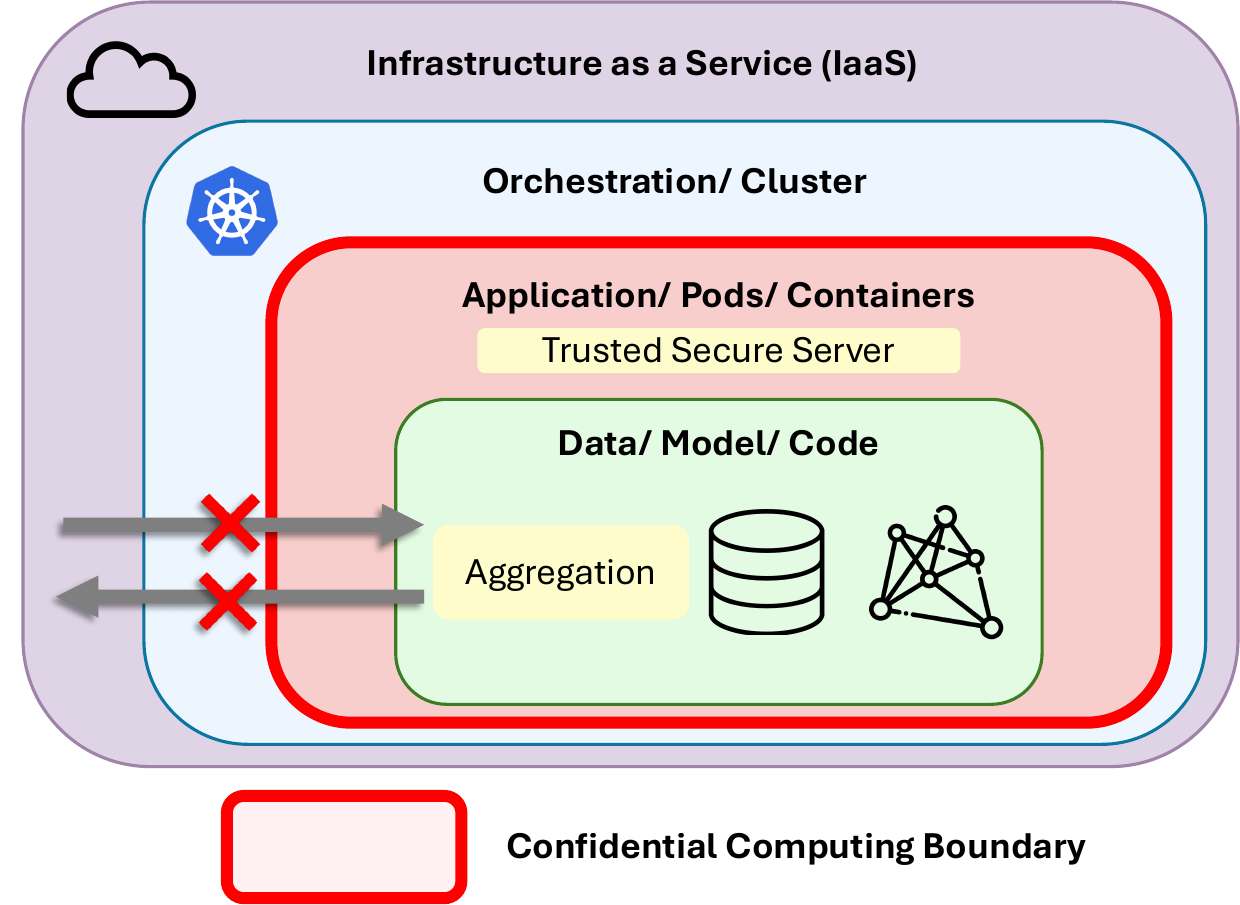}
    \end{center}
    \caption{\label{fig:threat-persona} Trusted Execution Environment Threat personas.}
\end{figure}

The hardware-based TEE protects aggregation with confidentiality and integrity guarantees against outer layers, including the infrastructure providers and orchestration tools shown in the threat personas in Figure \ref{fig:threat-persona}.

In addition, the system protects against threats within the server itself through memory encryption and remote attestation. The TEE keeps individual clients' updates encrypted even in memory. Remote attestation was originally designed to allow an enclave owner to verify the identity of a trusted binary executed in the cloud. In \pcs, clients can use the attestation quote to verify the binary's identity before sending updates. This design makes it impossible to update the trusted binary in the future without updating the clients simultaneously. To simplify the update process, verifiable logs can be used to record any changes made to the code that will run inside the enclave.

Together, TEE hosting and remote attestation meet the data and computation privacy goals in \sect{sec:threat_model_and_goals}.

\subsubsection{Defend Against Membership Attacks via Differential Privacy}
\label{sect:dp}


To address membership privacy from \sect{sec:threat_model_and_goals}, \pcs provides two Differential Privacy methods:

\textbf{Central DP at the enclave} (CDP). Applying DP at the central aggregator is the most straightforward and generalizable way to obtain DP. The aggregator adds noise to the aggregated results before releasing the federated model to clients or modelers.

\textbf{Local DP at the edge} (LDP). Local DP runs on each client. Each client noises local weights before uploading to the server, eliminating the need to trust the aggregator but typically reducing model utility.

\subsection{Reliability}
\label{sect:reliability}

For reliable operation under fragile rural infrastructure, \pcs uses multi-cluster orchestration for client and server cluster deployment, and supports both synchronous and asynchronous FL aggregation. The latter serves as a fallback to reduce the impact of stragglers and dropouts of client updates.

\begin{table*}[t]
    \centering
    \setlength{\tabcolsep}{4pt}
    \renewcommand{\arraystretch}{1.0}
    \begin{tabular}{|p{0.11\textwidth}|p{0.15\textwidth}|p{0.16\textwidth}|p{0.23\textwidth}|p{0.25\textwidth}|}
        \hline
        \textbf{Deployment} & \textbf{Data}
                            & \textbf{Model}
                            & \textbf{Training}
                            & \textbf{Privacy}                                                                                                                           \\
        \hline
        Nitrogen monitoring (NY)
                            & 3 farms (2 greenhouses, 1 field); 1{,}090 images.
                            & Leaf segmentation; U-Net + MobileNetV2 ($\alpha{=}0.35$).
                            & Pretrained on public dataset \cite{steininger2023cropandweed, hughes2015plantvillage}; fine-tuned on client data; aggregated with FedAdam.
                            & DP-FTRL Gaussian; central DP; per-layer adaptive clipping ($C_0{=}0.5$); R\'enyi-DP ($\delta{=}10^{-5}$).                                  \\
        \hline
        ET prediction (CA)
                            & 10 CIMIS stations of distinct climates; $34{,}107$ samples.
                            & Next-day ET regression; stacked LSTM ($h{=}64$, 2L).
                            & Trained on 2008--2020; tested on 2020--2025; aggregated with FedAvg.
                            & DP-FTRL Gaussian; central DP; global adaptive clipping ($C_0{=}0.5$); R\'enyi-DP ($\delta{=}10^{-5}$).                                     \\
        \hline
    \end{tabular}
    \caption{Summary of the two deployments on \pcs.}
    \label{tab:eval_setting}
\end{table*}

\subsubsection{Multi-cluster Orchestration on Farms}
\pcs models each farm participating in the system as a Kubernetes \cite{kubernetes} cluster. With Digital Agriculture, farms may have a variety of devices that collect, process or act on data, and a cluster is the most convenient way to work with these heterogeneous devices. The FL client binary can be deployed as a container on the farm cluster to make execution and deployment a managed task, without specialized personnel handling deployment operations on the farm. To scale to multiple farms, however, \pcs requires a multi-cluster orchestration framework that can deploy workloads and monitor deployment state from a single interface (Figure \ref{fig:design}). \pcs is general in its application, and FL framework changes can be made without needing changes to cluster configuration and vice versa. With clusters and containers, \pcs abstracts away the management of the growers' hardware, making it easier for the modeler to run Federated Learning workloads.

For a distributed system to run on the farm, it needs to withstand network partitions for extended periods of time \cite{trustinfood2021farmerdata}. With a multi-cluster scenario, a farm cluster can actively run without being connected to the control plane for the multi-cluster system; as long as there is a valid deployment on the cluster during the partition, the cluster will continue to run. For a multi-cluster control plane that allows for disconnected operations, \pcs uses KubeStellar, an open-source multi-cluster orchestration and deployment framework \cite{kubestellar}. KubeStellar uses a hub-and-spoke model, where a single hub running KubeStellar can allow farm clusters to enroll in the network to accept and deploy Kubernetes resources. The architecture is detailed in \sect{sec::implementation:orchestration}.

\subsubsection{Dynamic Adaptation of Asynchronous FL}

The coordinator (\sect{sec:TEE-based_TSS}) monitors server-client communication. The aggregator that starts in synchronous FL switches to asynchronous FL when the coordinator detects stalls from stragglers or dropouts, often caused by network issues or client outages.

%% file: implementation.tex
\section{Implementation}
\label{sec::implementation}

\subsection{Orchestration}
\label{sec::implementation:orchestration}

To manage multiple farm clusters from a single interface, we need a performant and
scalable solution that works well in a rural environment. \pcs uses KubeStellar, an open-source multi-cluster orchestration framework to orchestrate and deploy workloads down to
remote clusters. The KubeStellar architecture contains two important components: The
\textbf{Workload Definition Space (WDS)} which accepts manifests to be deployed down to
spoke clusters, and the \textbf{Inventory and Transport Space (ITS)}, that communicates
with spoke clusters, aggregates status and reports, and deploys any objects defined in
the WDS downstream. Before defining a workload, users can define a \verb|BindingPolicy| object that matches workloads to clusters using resource labels. Since each
farm is an independent Kubernetes cluster, it does not need a persistent connection to
the KubeStellar hub to function. On disconnection, any new workload updates defined with
KubeStellar are held in stasis until the cluster reconnects. The farm cluster also
continues to function offline, where applicable, while it is disconnected. Unlike a
traditional Kubernetes pod, the cluster is not replaced with a new deployment once it reconnects.

\subsection{TEE-based Trusted Secure Server}
\label{sec::implementation:tee}

Confidential VMs (CVMs) allow the execution of unmodified binaries within a TEE \cite{confidentialVMsExplained}, where both the code and data are protected from the Virtual Machine Monitor (VMM) and the host. CVMs are supported by VM-level ISA extensions such as Intel TDX \cite{intelTDX}, AMD SEV-SNP \cite{sev2020strengthening}, and more recently Arm CCA \cite{armCCA}, and are available on major cloud vendor platforms.

To run \pcs aggregation in a multi-cluster deployment, we integrate the open-source Confidential Containers project (CoCo), from the Cloud Native Computing Foundation, within our framework to deploy TEEs on demand on the public cloud \cite{confidentialcontainers}. CoCo is a cloud-native tool that can allow us to provision and deploy aggregators for every FL task for \pcs, and it is built on top of Kata Containers, a container runtime that provides hardware virtualization for containers by running them within lightweight VMs \cite{katacontainers}.

With CoCo, the aggregation workload runs inside a CVM, and the server application can run on trusted hardware without any modifications. The state of the VM and the identity of the workload can be remotely attested, which is a key step before clients share any information with the TSS. There are studies that discuss CVM performance overheads \cite{confidentialVMsExplained}, including significant boot times and memory acceptance latency, but we find that those effects are amortized by having our trusted secure aggregator long-running.




\subsection{Client}

The client binary for \pcs uses the gRPC protocol to communicate with the TSS. To participate in the FL training process, it pulls the model from the remote server and trains it on locally available data. The client also collects data in a single location for cases that require accumulating data from distributed sources, and can spawn multiple training tasks to be executed at the same time. The client also reports metrics such as training loss and time elapsed to the server for ongoing training tasks.

%% file: evaluation.tex
\begin{figure*}[th]
    \centering
    \captionsetup[subfigure]{skip=0pt}

    \begin{subfigure}[b]{\textwidth}
        \centering
        \includegraphics[width=\textwidth]{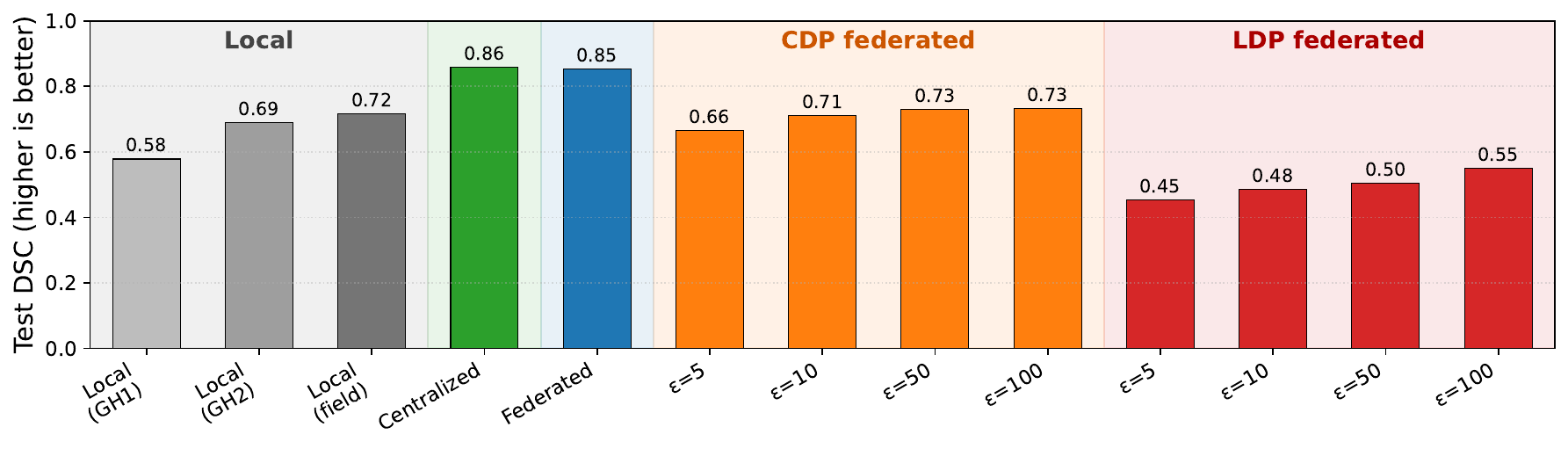}
        \caption{Segmentation Accuracy Score, \DSC, across training settings and privacy budgets, higher is better.}
        \label{fig:tomato-dice}
    \end{subfigure}
    \vspace{-1em}

    \begin{subfigure}[b]{\textwidth}
        \centering
        \includegraphics[width=\textwidth]{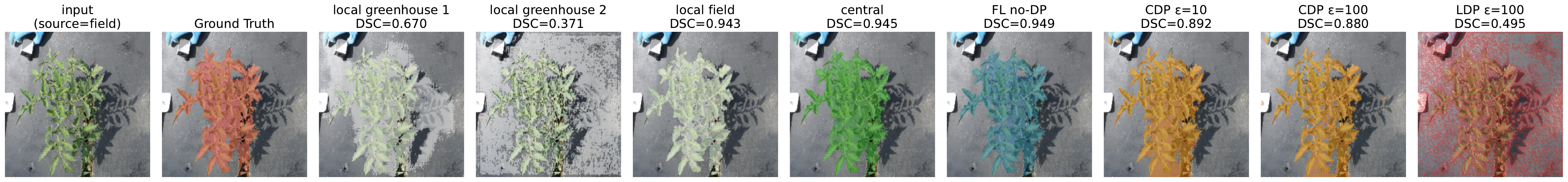}
        \caption{Segmentation predictions on one test image of models trained under different settings.}
        \label{fig:tomato-segmentation}
    \end{subfigure}

    \caption{\textbf{\TomatoDeployment}: Segmentation accuracy and visualization (see \sect{sec::eval-accuracy} for details). The color configuration of bars in \ref{fig:tomato-dice} matches the color configuration of masks in \ref{fig:tomato-segmentation}, except for the local models that remain neutral such that the background stays visible.}
    \label{fig:tomato-comparison}
\end{figure*}

\section{Deployment}\label{sec::evaluation}




For the deployment setup, we run the server on an AMD EPYC 7R13 Processor (4 vCPUs) and 16GB of memory in the public cloud. The farm clusters each run on a Raspberry Pi Model 5 (8GB) or Model 4 (8GB) with a lightweight Kubernetes distribution \cite{k3s}. The questions that were evaluated via the deployment include:
\begin{itemize}[itemsep=1pt, topsep=2pt, parsep=0pt, partopsep=0pt]
    \item Can the system protect data privacy without sacrificing model accuracy? (\sect{sec::eval-accuracy})
    \item How does the framework hold up under farm client dropouts? (\sect{sec::eval-async})
    \item What overhead does TEE-secured aggregation add at the aggregator/cloud-based server and end-to-end performance? (\sect{sec::eval-tee})
\end{itemize}

Table~\ref{tab:eval_setting} summarizes the setup for both deployments. This section uses three metrics to evaluate accuracy and privacy:
\\
\textbf{\DSC} for segmentation accuracy on \TomatoDeployment. \DSC measures the overlap between the predicted and ground-truth segmentation masks \cite{zou2004statistical}. Values above 0.70 are generally considered good for semantic segmentation \cite{zou2004statistical}.
\\
$\mathbf{R^2}$ \textbf{score} for prediction accuracy on \ETDeployment. $R^2$ is the fraction of variance in actual ET explained by the model \cite{r_squre}. Values of 0.8 or higher are considered good for regression in scientific applications.
\\
\textbf{Privacy budget $\mathbf{\epsilon}$} is the core parameter of $(\epsilon, \delta)$-Differential Privacy \cite{dp}. $\epsilon$ bounds how much an attacker can distinguish two datasets that differ by one record (one daily sample or one training image). Larger $\epsilon$ permits less noise and higher utility. Values from 1 to 10 are common in practice \cite{ponomareva2023howtodpfy}.
Privacy composition across rounds is tracked by the R\'enyi-DP accountant \cite{mironov2017renyi}. Both deployments add noise using DP-FTRL \cite{kairouz2021practical}, a binary-tree mechanism that reduces cumulative noise across rounds.

\begin{figure*}[t]
    \centering
    \includegraphics[width=\textwidth]{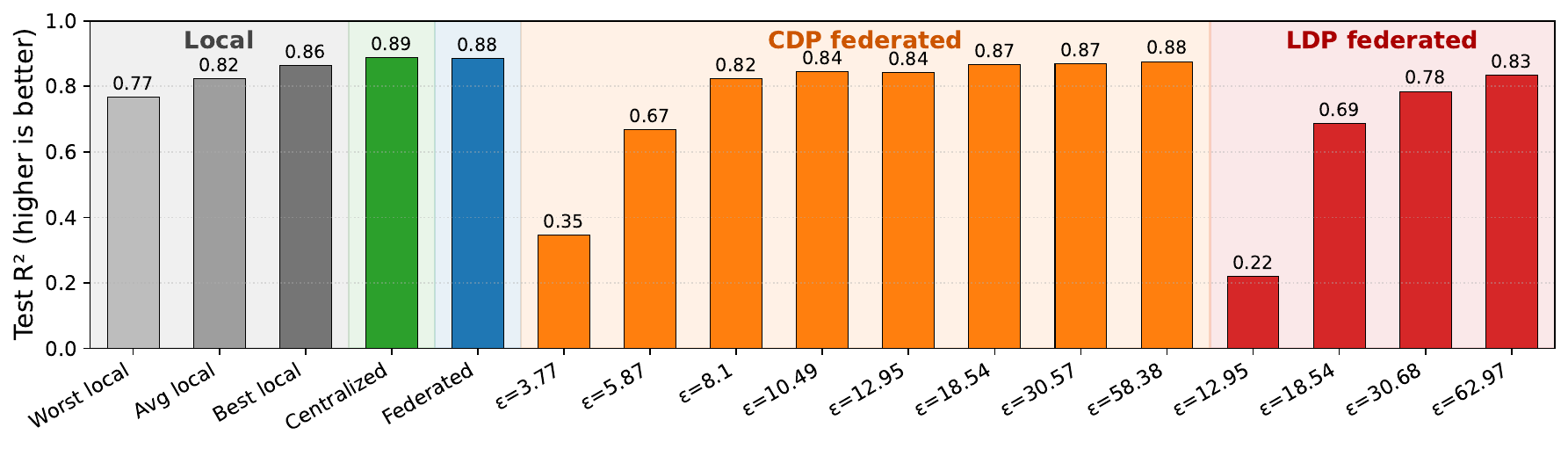}
    \caption{\textbf{\ETDeployment}: Prediction accuracy, $R^2$, across training settings and privacy budgets. See $\sect{sec::eval-accuracy}$ for details.}
    \label{fig:et-r2}
\end{figure*}

\subsection{Accuracy and Privacy Tradeoffs}
\label{sec::eval-accuracy}

\pcs uses two privacy mechanisms with accuracy costs: Federated Learning, which keeps raw data on the client, and Differential Privacy, which adds calibrated noise. To quantify their impact, five training settings are compared per deployment, ordered from least to most privacy-preserving: \textit{Centralized} (all clients' raw data pooled at one trainer), \textit{Federated} (each client trains locally; only model updates are aggregated), \textit{Federated + CDP} (noise added at the secure enclave during aggregation), \textit{Federated + LDP} (noise added at each client before upload), and \textit{Local} (each client trains independently with no aggregation).

For the \TomatoDeployment, Figure~\ref{fig:tomato-dice} reports test \DSC across the five training settings.
Federated Learning achieves \DSC comparable to centralized training (\tomatoFL{} vs.\ \tomatoCentralized) while keeping raw images on-device, and improves on the best local baseline (field site, $0.72$) by $0.13$.
This is because the three client sites differ in light conditions, plant sizes, and camera angles, and cross-client aggregation lets the model learn from this heterogeneity despite the limited per-site image volume.
Adding DP reduces accuracy differently for the two placements:
CDP saturates at \DSC \tomatoCDP, matching the best local baseline once $\epsilon \geq \tomatoCDPEPS$,
while LDP fails to converge even at the loosest budget evaluated; at $\epsilon = 100$, LDP achieves only $0.55$, below the worst local baseline ($0.58$).

Figure~\ref{fig:tomato-segmentation} visualizes one field image under each setting: FL without DP produces a leaf mask visually indistinguishable from centralized training; CDP at moderate $\epsilon$ still retains most of the leaf area, whereas LDP generates a noisy mask.

\textbf{For this fine-tuning workload}, Federated Learning without DP achieves near-centralized accuracy and exceeds every local baseline; CDP is the viable privacy mechanism when DP is required, while LDP underperforms single-site training.

For the \ETDeployment, Figure~\ref{fig:et-r2} reports test $R^2$ across the five training settings. Federated Learning matches centralized training ($R^2$ \etFL{} vs.\ \etCentralized) while keeping raw station data on-device, and exceeds every local baseline (best $0.86$, average $0.82$, worst $0.77$).
The ten stations span distinct climate zones with varying historical record lengths, so the federated model benefits from learning across this heterogeneity, and delivers strong $R^2$ at every station.

Adding DP reduces accuracy differently for the two placements: under CDP, $R^2$ reaches \etCDP at $\epsilon = \etCDPEPS$, then plateaus close to the FL baseline. LDP reaches a competitive $R^2$ ($0.83$) at a much larger $\epsilon \approx 63$.

\begin{figure*}[t]
    \centering
    \captionsetup[subfigure]{skip=2pt}

    \begin{subfigure}[b]{0.32\textwidth}
        \centering
        \includegraphics[width=\linewidth]{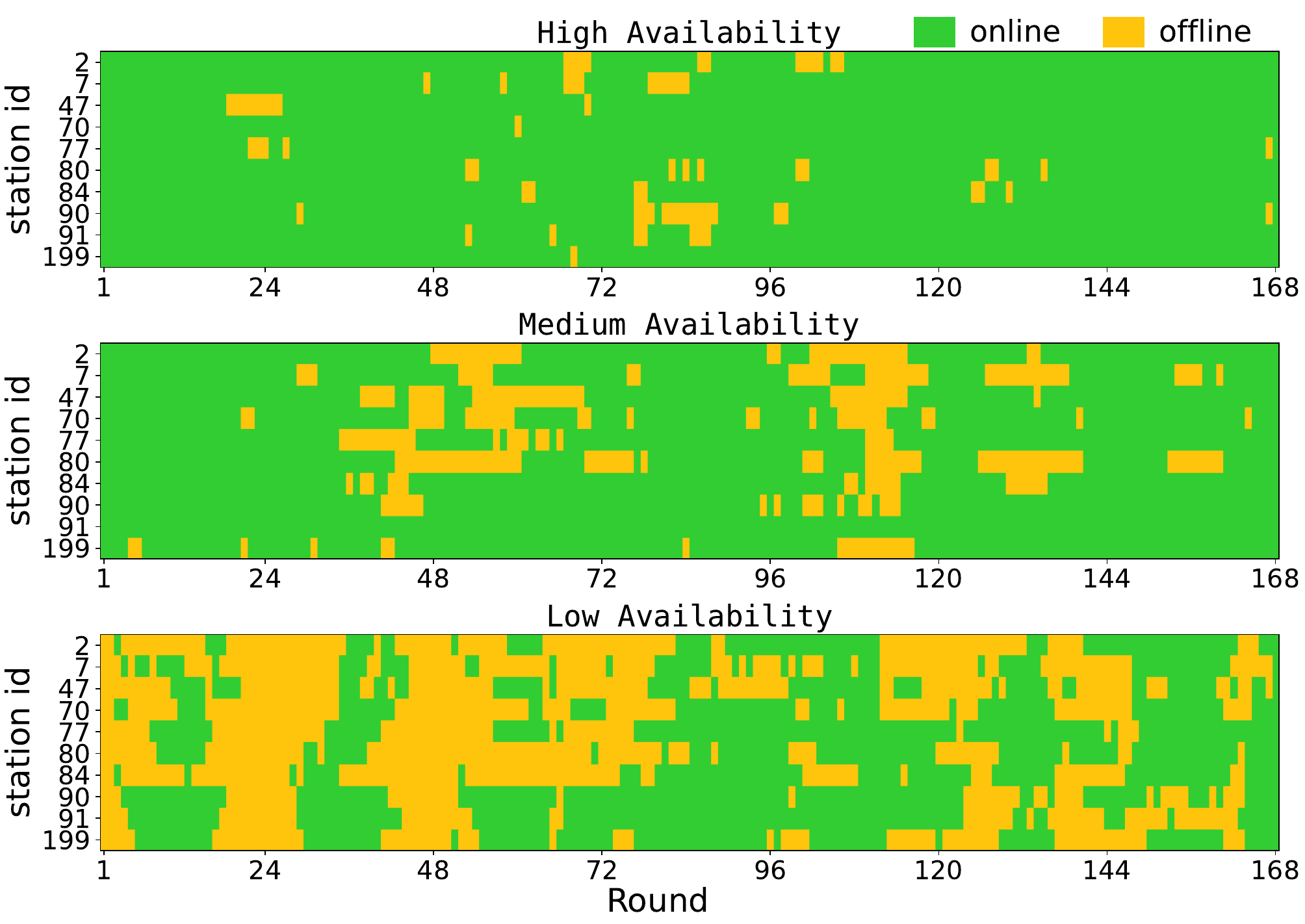}
        \caption{Three one-week availability traces from 10 stations under high, medium, low availability.}
        \label{fig:et-avail-heatmap}
    \end{subfigure}
    \hfill
    \begin{subfigure}[b]{0.32\textwidth}
        \centering
        \includegraphics[width=\linewidth]{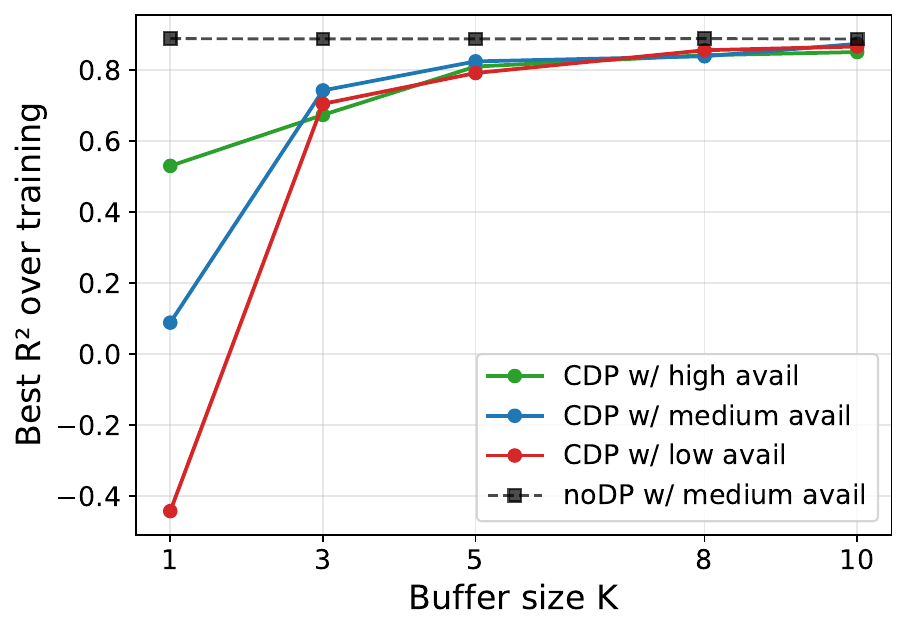}
        \caption{Best $R^2$ vs.\ buffer size $K$ per availability, higher is better.}
        \label{fig:et_async_best_r2}
    \end{subfigure}
    \hfill
    \begin{subfigure}[b]{0.32\textwidth}
        \centering
        \includegraphics[width=\linewidth]{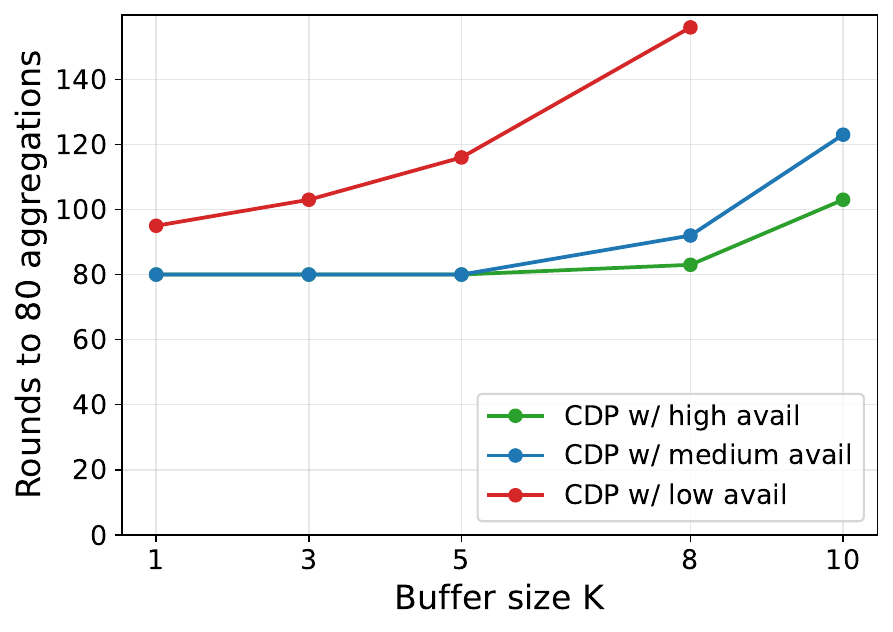}
        \caption{Rounds vs.\ buffer size $K$ per availability, lower is better.}
        \label{fig:et_async_rounds}
    \end{subfigure}
    \caption{\textbf{\ETDeployment}: AsyncFL (FedBuff) evaluation with trace-replayed CIMIS station availability. See $\sect{sec::eval-async}$ for details. Buffer size $K$ is the number of client updates the server waits for before aggregating.}
    \label{fig:et_async}
\end{figure*}

\begin{figure}[t]
    \begin{center}
        \includegraphics[width=\linewidth]{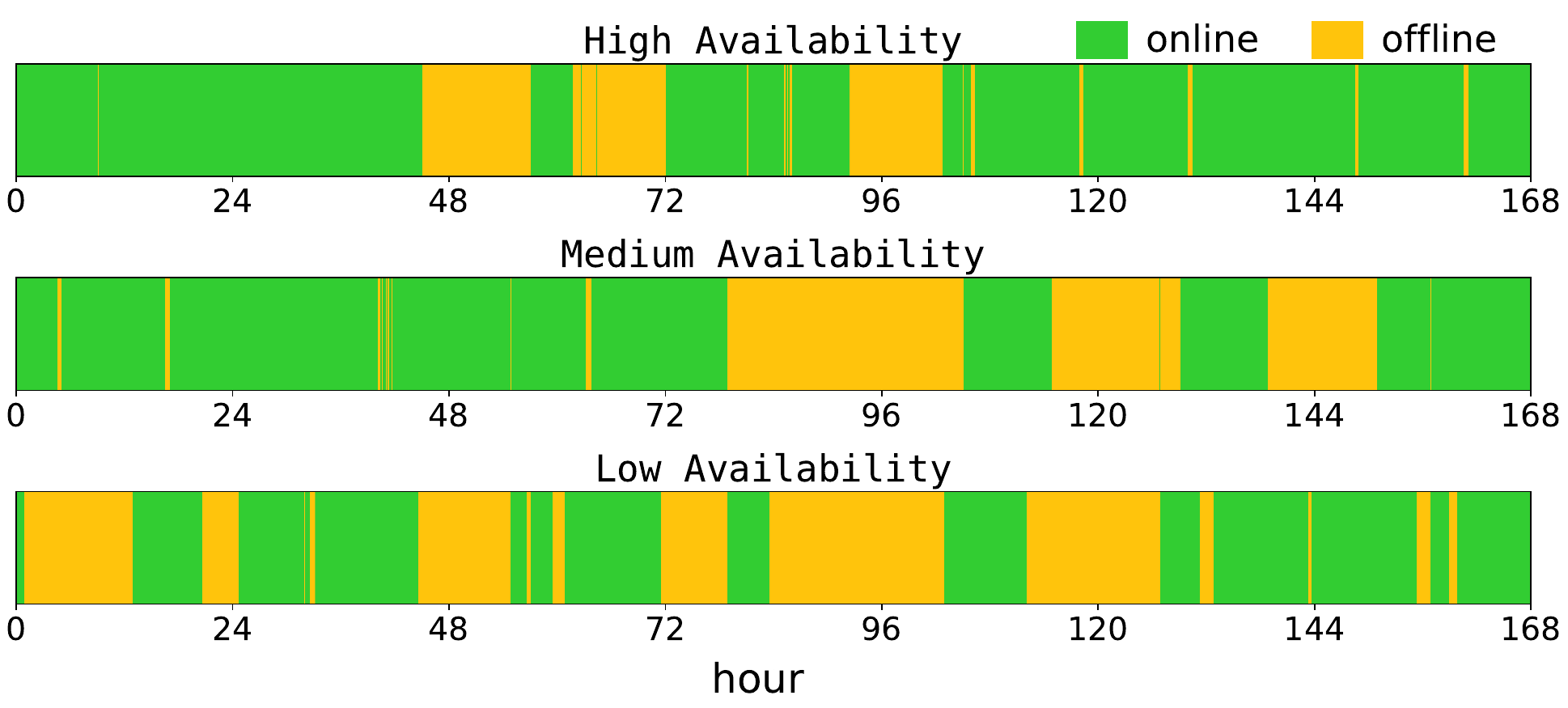}
    \end{center}
    \caption{\label{fig:tomato-avail-trace} \textbf{\TomatoDeployment}: Three representative one-week availability traces from a greenhouse client under high, medium, and low availability.}
\end{figure}

\textbf{For this regression workload}, Federated Learning without DP achieves near-centralized accuracy and uniformly exceeds local training; both DP variants can reach near-centralized $R^2$, but CDP does so at moderate $\epsilon$ while LDP requires a much larger budget.

\textbf{Takeaway}: FL with Central DP is the recommended privacy-preserving configuration: it improves the worst single-site model performance (by 22.4\% on Nitrogen, 9.1\% on ET) while keeping raw data on-device, and is more $\epsilon$-efficient than Local DP for membership privacy.

\subsection{Async FL Robustness}
\label{sec::eval-async}

To evaluate AsyncFL (FedBuff) robustness under client dropouts, \pcs replays real availability traces from both deployments and sweeps the buffer size $K$, the number of client updates the server waits for before each aggregation.

For the \TomatoDeployment, Figure~\ref{fig:tomato-avail-trace} shows three representative one-week availability traces from a greenhouse client: high (client-uptime 79.1\%), medium (67.3\%), and low (55.6\%) availability. \pcs replays these traces in the $K$-sweep experiment. With only 3 clients, CDP requires $K = 3$ to converge above \DSC 0.7 because noise must be averaged across all clients; smaller $K$ leaves the noise unaveraged and CDP fails to converge.

For the \ETDeployment, Figure~\ref{fig:et-avail-heatmap} shows three representative one-week windows captured from the ten-month operational deployment of the ten CIMIS stations: high (station-uptime 95.4\%), medium (83.5\%), and low (52.3\%) availability. Each row of a panel is a station and each column is one aggregation round (one hour of wall-clock time). \pcs sweeps $K$ from 1 to 10 under CDP, measuring accuracy and the rounds needed to complete 80 aggregations.

Figure~\ref{fig:et_async_best_r2} shows the model accuracy ($R^2$) as $K$ varies across the three availabilities. When $K$ is small ($K \leq 3$) with CDP, the model collapses because the aggregator cannot average the noisy updates with a small number of clients. The accuracy is worse with lower availability. When $K$ is large enough to average noise across clients ($K \geq 8$), $R^2$ reaches 0.85--0.87 across all three tiers, essentially matching the no-DP baseline ($R^2 = 0.88$). Figure~\ref{fig:et_async_rounds} shows the number of rounds needed to complete 80 aggregations. Under medium or high availability, $K \leq 5$ completes 80 aggregations in 80 rounds; under low availability, larger $K$ slows progress ($K=8$ needs 156 rounds, and $K=10$ fails to complete within the 168-round window). Together, $K = 8$ is the operational sweet spot: it brings $R^2$ to within $0.03$ of the no-DP baseline across all three availability tiers, while remaining the largest $K$ that completes 80 aggregations within 168 rounds with all three availabilities.

\textbf{Takeaway}: With workload-tuned buffer size $K$, FedBuff matches the synchronous baseline accuracy under real availability traces, serving as a reliable fallback when client availability degrades.

\subsection{TEE Overhead}
\label{sec::eval-tee}


\begin{table}[t]
    \centering
    \small
    \setlength{\tabcolsep}{7pt}
    \begin{tabular}{@{}llccr@{}}
        \toprule
         & $E$ & Bare (s) & TEE (s) & $\Delta$ \\
        \midrule
        \multirow{3}{*}{Nitrogen}
         & 1   & 146.07   & 146.01  & $+0.0\%$ \\
         & 2   & 289.87   & 286.61  & $-1.1\%$ \\
         & 4   & 574.97   & 573.00  & $-0.3\%$ \\
        \midrule
        \multirow{3}{*}{ET}
         & 1   & 14.40    & 14.49   & $+0.6\%$ \\
         & 2   & 26.31    & 26.36   & $+0.2\%$ \\
         & 4   & 48.98    & 49.33   & $+0.7\%$ \\
        \bottomrule
    \end{tabular}
    \caption{Median per-round time for bare and TEE server configurations across local epoch $E$.
        $\Delta$ is relative to bare.}
    \label{tab:tee-overhead}
\end{table}

From prior work, using TEEs has been known to introduce a measurable latency overhead \cite{confidentialVMsExplained}. With \pcs, however, this overhead becomes less significant as more time is spent training the model on the client.
To measure this overhead, \pcs runs the server in two configurations: a vanilla Kubernetes pod (bare) and a confidential peer pod (TEE).
Clients run on Raspberry Pi devices throughout.
For both deployments, the local training epoch size $E$ (the number of passes over the local dataset per round) is varied from 1 to 4 to scale client-side training time, which grows linearly with $E$.
Table \ref{tab:tee-overhead} reports median per-round times, and shows that the overhead is within \TEEOverheadBound across both deployments and all epoch sizes.
Small negative values on the \TomatoDeployment are within measurement noise. \pcs amortizes the TEE overhead on the server through client-side training time.

\textbf{Takeaway}:
TEE-secured aggregation adds less than \TEEOverheadBound\ per-round latency overhead across both deployments, amortized by client-side training time on Raspberry Pi hardware.

%% file: discussion.tex
\section{Limitations}

%
The \pcs deployment has several limitations.


\textbf{Honest-but-curious threat model.} \pcs assumes an honest-but-curious threat model, following the conventional assumption that fits the closed collaborator pool of agricultural research deployments. Malicious behaviors such as data poisoning and model poisoning are out of scope.


\textbf{TEE memory exploits.} \pcs secures aggregation via TEE, implemented as Confidential Containers \cite{confidentialcontainers}. The security of Confidential Containers, however, is contingent upon the underlying secure hardware; recent 100\% software-only exploits such as BreakFAST \cite{giersfeld2026breakfast}, Fabricked \cite{schlueter2026fabricked} and Staleus \cite{schlueter2026staleus}, for AMD SEV-SNP, breach TEE security and can limit protection afforded by the framework. It is at the infrastructure providers' discretion to apply the latest security updates to mitigate these security vulnerabilities.   

\textbf{Async FL bias.} FedBuff can theoretically bias the aggregated model toward clients with higher availability under non-IID data. Our evaluation shows this does not affect final accuracy in either deployment, but the result may not generalize to deployments with greater availability disparity.

%% file: related-work.tex
\section{Related Work}
\label{sect::related}

Digital Agriculture is revolutionizing farm management through data-driven techniques, offering significant financial, environmental, and societal benefits \cite{Mueller2012, https://doi.org/10.1002/jsfa.9346}. DA involves collecting and processing vast amounts of data to provide farmers with actionable insights \cite{6679872, vasisht2017farmbeats}. Importantly, privacy is a key concern for adoption \cite{trustinfood2021farmerdata}.
\subsection{Machine Learning for Digital Agriculture}
\label{subsect::DA-in-pathology}

A 2025 survey by Dembani et al. \cite{dembani2025agricultural} investigates the current use of Federated Learning in agriculture to solve grower privacy concerns, and discusses successful applications across a variety of use cases, including pest and disease detection, crop yield prediction, and precision resource management. It reports effective, privacy-preserving and close-to-baseline results for predicting crop yield for soybean and maize with FL using FedAvg and ResNet ML models. The survey shows, however, that scalability and computational resources are ongoing technical challenges that need to be solved before scaling to a large number of farms for FL. This alludes to the development of a resource-aware FL framework, such as \pcs, that can work around cluster resources and availability to let more farms participate in the framework.

\subsection{Platform Orchestration for Digital Agriculture}
\label{subsect::platform-orch-DA}
Deploying and managing infrastructure on the farm is a non-trivial task. Platform orchestration for Digital Agriculture coordinates various components and resources to run applications across distributed environments.\Comment{It integrates data collection, storage, processing, and analysis tools to support farm decision-making.} Work done by Rubambiza et al. \cite{MLpipeline} demonstrates the use of cloud-based solutions for detecting grape disease in vineyards in the U.S. With \pcs, we build upon their work to introduce privacy guarantees for growers consuming such ML models on their farms. As agricultural data grows in volume and complexity, we are interested in protecting the right to privacy for people who own that data.



Another feature of an agricultural setting is fragile network connectivity of farms in rural locations \cite{trustinfood2021farmerdata}. Consequently, DA platforms must be designed to function even when network connectivity is unavailable. The concept of disconnected operation, as demonstrated in systems like Coda and the Bayou storage system, enables clients to access data during temporary failures of a shared data repository \cite{coda, bayou}. Coda enhances availability through client-side caching \cite{coda}. Bayou incorporates application-specific mechanisms for conflict detection and resolution \cite{bayou}. Similarly, the workload running on the farm with \pcs is designed to continue to function offline, and to fetch new workloads or updates from the central orchestrator once reconnected.

\subsection{Privacy Preservation in Machine Learning}

Open-source FL frameworks such as Flower \cite{flower, flower-dp}, FedML \cite{fedml}, and NVIDIA FLARE \cite{nvflare-cc} provide general-purpose FL infrastructure.
However, none of these offers specialized configurations to deal with the privacy requirements of an agricultural deployment. In particular, neither Flower nor FedML provide buffered asynchronous aggregation (FedBuff \cite{nguyen2022federatedlearningbufferedasynchronous}) to tolerate fragile rural infrastructure or TEE-secured aggregation in the cloud.
  Although FLARE supports TEE-secured aggregation via confidential computing \cite{nvflare-cc}, it requires TEE-capable hardware at every participant, whereas \pcs{} confines the TEE to the cloud aggregator so farm clients run on commodity Raspberry Pis.
  All three leave client provisioning and partition recovery to specialized personnel; \pcs{} instead manages farm clusters through a multi-cluster orchestration interface (\sect{sect:reliability}).

  Domain-specialized and industrial FL systems demonstrate that such gaps are best closed by tailoring frameworks to their setting.
  Fed-BioMed \cite{Cremonesi2025} recognizes the mismatch between generalized FL frameworks and the privacy requirements of healthcare, and describes a system that satisfies those constraints.
  At industrial scale, GoogleFL \cite{bonawitz2019googlefl}, AFL (Apple) \cite{paulik2021applefl}, and Papaya FL (Meta) \cite{huba2022papaya} demonstrate the practical applicability of privacy-preserving FL in large-scale consumer settings.

  Our work draws inspiration from these frameworks and applies their findings to agriculture scenarios.
  In particular, Papaya describes synchronous and asynchronous FL strategies for scaling to millions of devices, with a Trusted Secure Aggregator that processes user updates off-device using \textit{random masking} for privacy \cite{huba2022papaya}.
  \pcs{} demonstrates FL in an agricultural setting with Differential Privacy instead: DP incurs an acceptable accuracy cost while providing protection against membership inference attacks, which random masking does not \cite{shokri2017membership}.
  Additionally, \pcs implements an open-source, vendor-agnostic framework; users can deploy it on-premises or on the public cloud of their choosing. The platform can be used without vendor lock-in, and may also be generalized for other agricultural workloads because of the flexibility afforded by CVMs.


%% file: conclusion.tex
\section{Conclusion}
\label{sec::conclusion}
Artificial Intelligence (AI) has great potential to improve agricultural practices~\cite{unfss2_highlights}, but data privacy is one of the most critical requirements that need to be satisfied before farmers adopt AI \cite{trustinfood2021farmerdata}. In this work, we take a step in deploying and evaluating improved data privacy approaches in agriculture. Specifically, we report on our experience working in collaboration with growers in two deployments; one monitoring nitrogen \Comment{levels via sentinel plants }in New York for \TomatodeployedTenure and a second predicting evapotranspiration in California for \ETdeployedTenure. With the \pcslong, growers retain control over their privacy when participating in the training of Machine Learning models to assist them with resource allocation decisions. \pcs uses a Kubernetes-based multi-cluster Federated Learning system with Differential Privacy that supports both synchronous and asynchronous FL training algorithms. We show that \pcs is resilient by design; both the multi-cluster orchestration framework and the asynchronous fallback protocols allow farm clusters to participate in the FL system with fragile rural infrastructure.
\pcs improves the worst single-site model accuracy by 22.4\% for nitrogen monitoring and 9.1\% on evapotranspiration prediction, showing that \pcs improves model utility from single-site training while preserving privacy.
For future work, we intend to extend \pcs to include more agricultural use cases, and scale the system up to a national infrastructure-level deployment.

%% file: ref.bib
@article{Almarshadi2011EffectsOP,
  author  = {Almarshadi, Mohammed and Ismail, Saleh},
  year    = {2011},
  month   = {03},
  pages   = {},
  title   = {Effects of Precision Irrigation on Productivity and Water Use Efficiency of Alfalfa under Different Irrigation Methods in Arid Climates},
  volume  = {7},
  journal = {Journal of Applied Sciences Research}
}

@article{Kim2009SoilMS,
  title   = {Soil macronutrient sensing for precision agriculture.},
  author  = {Hak Jin Kim and Kenneth A. Sudduth and John Hummel},
  journal = {Journal of environmental monitoring : JEM},
  year    = {2009},
  volume  = {11 10},
  pages   = {
             1810-24
             },
  url     = {https://api.semanticscholar.org/CorpusID:6194919}
}

@article{altman2024techbrief,
  author    = {Altman, Micah and Cohen, Aloni and Nissim, Kobbi},
  title     = {{ACM TechBrief: Data Privacy Protection}},
  year      = {2024},
  isbn      = {979-8-4007-1115-2},
  publisher = {Association for Computing Machinery},
  address   = {New York, NY, USA}
}

@article{gold2021plant,
  title     = {Plant Disease Sensing: Studying Plant-Pathogen Interactions at Scale},
  author    = {Gold, Karen M.},
  journal   = {mSystems},
  volume    = {6},
  number    = {6},
  pages     = {e01228-21},
  year      = {2021},
  publisher = {American Society for Microbiology},
  doi       = {10.1128/mSystems.01228-21},
  url       = {https://doi.org/10.1128/mSystems.01228-21}
}

@misc{space2farmNews,
  author = {Krishna Ramanujan},
  title  = {From space to farm: readying NASA satellites to help growers},
  year   = {2024},
  month  = aug,
  url    = {https://news.cornell.edu/stories/2024/08/space-farm-readying-nasa-satellites-help-growers},
  note   = {Cornell Chronicle, August 22, 2024}
}

@article{MLpipeline,
  author  = {Rubambiza, Gloire and Galvan, Fernando and Pavlick, Ryan and Weatherspoon, Hakim and Gold, Kaitlin},
  year    = {2023},
  month   = {05},
  pages   = {},
  title   = {Toward Cloud‐Native, Machine Learning Base Detection of Crop Disease With Imaging Spectroscopy},
  volume  = {128},
  journal = {Journal of Geophysical Research: Biogeosciences},
  doi     = {10.1029/2022JG007342}
}

@article{MLMethods,
  author  = {Jez, Joseph M. and Topp, Christopher N. and Silva, Gonçalo and Tomlinson, Jenny and Onkokesung, Nawaporn and Sommer, Sarah and Mrisho, Latifa and Legg, James and Adams, Ian P. and Gutierrez-Vazquez, Yaiza and Howard, Thomas P. and Laverick, Alex and Hossain, Oindrila and Wei, Qingshan and Gold, Kaitlin M. and Boonham, Neil},
  title   = {{Plant pest surveillance: from satellites to molecules}},
  journal = {Emerging Topics in Life Sciences},
  volume  = {5},
  number  = {2},
  pages   = {275-287},
  year    = {2021},
  month   = {03},
  issn    = {2397-8554},
  doi     = {10.1042/ETLS20200300},
  url     = {https://doi.org/10.1042/ETLS20200300},
  eprint  = {https://portlandpress.com/emergtoplifesci/article-pdf/5/2/275/912777/etls-2020-0300c.pdf}
}

@inproceedings{comosum,
  author    = {Gloire Rubambiza and Shiang-Wan Chin and Mueed Rehman and Sachille Atapattu and Jos{\'e} F. Mart{\'\i}nez and Hakim Weatherspoon},
  title     = {Comosum: An Extensible, Reconfigurable, and {Fault-Tolerant} {IoT} Platform for Digital Agriculture},
  booktitle = {2023 USENIX Annual Technical Conference (USENIX ATC 23)},
  year      = {2023},
  isbn      = {978-1-939133-35-9},
  address   = {Boston, MA},
  pages     = {197--214},
  url       = {https://www.usenix.org/conference/atc23/presentation/rubambiza},
  publisher = {USENIX Association},
  month     = jul
}

@misc{kubernetes,
  title  = {Kubernetes},
  author = {{Kubernetes}},
  year   = {2024},
  url    = {https://kubernetes.io/},
  note   = {Accessed: 2024-09-28}
}

@misc{kubestellar,
  title  = {KubeStellar Release 0.24.0 Readme},
  author = {{KubeStellar}},
  year   = {2024},
  url    = {https://docs.kubestellar.io/release-0.24.0/readme/},
  note   = {Accessed: 2024-09-28}
}

@article{coda,
  author     = {Kistler, James J. and Satyanarayanan, M.},
  title      = {Disconnected operation in the Coda File System},
  year       = {1992},
  issue_date = {Feb. 1992},
  publisher  = {Association for Computing Machinery},
  address    = {New York, NY, USA},
  volume     = {10},
  number     = {1},
  issn       = {0734-2071},
  url        = {https://doi.org/10.1145/146941.146942},
  doi        = {10.1145/146941.146942},
  journal    = {ACM Trans. Comput. Syst.},
  month      = feb,
  pages      = {3–25},
  numpages   = {23}
}

@article{bayou,
  author     = {Terry, D. B. and Theimer, M. M. and Petersen, Karin and Demers, A. J. and Spreitzer, M. J. and Hauser, C. H.},
  title      = {Managing update conflicts in Bayou, a weakly connected replicated storage system},
  year       = {1995},
  issue_date = {Dec. 3, 1995},
  publisher  = {Association for Computing Machinery},
  address    = {New York, NY, USA},
  volume     = {29},
  number     = {5},
  issn       = {0163-5980},
  url        = {https://doi.org/10.1145/224057.224070},
  doi        = {10.1145/224057.224070},
  journal    = {SIGOPS Oper. Syst. Rev.},
  month      = dec,
  pages      = {172–182},
  numpages   = {11}
}

@article{Mueller2012,
  author  = {Nathaniel D. Mueller and James S. Gerber and Matt Johnston and Deepak K. Ray and Navin Ramankutty and Jonathan A. Foley},
  title   = {Closing yield gaps through nutrient and water management},
  journal = {Nature},
  volume  = {490},
  number  = {7419},
  pages   = {254--257},
  year    = {2012},
  doi     = {10.1038/nature11420},
  url     = {https://doi.org/10.1038/nature11420}
}

@article{https://doi.org/10.1002/jsfa.9346,
  author   = {Shepherd, Mark and Turner, James A and Small, Bruce and Wheeler, David},
  title    = {Priorities for science to overcome hurdles thwarting the full promise of the ‘digital agriculture’ revolution},
  journal  = {Journal of the Science of Food and Agriculture},
  volume   = {100},
  number   = {14},
  pages    = {5083-5092},
  doi      = {https://doi.org/10.1002/jsfa.9346},
  url      = {https://scijournals.onlinelibrary.wiley.com/doi/abs/10.1002/jsfa.9346},
  eprint   = {https://scijournals.onlinelibrary.wiley.com/doi/pdf/10.1002/jsfa.9346},
  year     = {2020}
}

@inproceedings{6679872,
  author    = {Tso, Fung Po and White, David R. and Jouet, Simon and Singer, Jeremy and Pezaros, Dimitrios P.},
  booktitle = {2013 IEEE 33rd International Conference on Distributed Computing Systems Workshops},
  title     = {The Glasgow Raspberry Pi Cloud: A Scale Model for Cloud Computing Infrastructures},
  year      = {2013},
  volume    = {},
  number    = {},
  pages     = {108-112},
  doi       = {10.1109/ICDCSW.2013.25}
}

@misc{precisionAgERS,
  author       = {Jonathan McFadden and Katherine Lim},
  title        = {Precision agriculture use increases with farm size and varies widely by technology},
  year         = 2024,
  howpublished = {\url{https://www.ers.usda.gov/data-products/charts-of-note/chart-detail?chartId=110550}},
  note         = {Accessed: 2025-10-23}
}

@misc{precisionAgGAO,
  author       = {{United States Government Accountability Office}},
  title        = {Precision Agriculture: Benefits and Challenges for Technology Adoption and Use},
  year         = 2024,
  howpublished = {\url{https://www.gao.gov/products/gao-24-105962}},
  note         = {Accessed: 2025-10-23}
}

@inproceedings{mycelium,
  title     = {Mycelium: Large-scale distributed graph queries with differential privacy},
  author    = {Roth, Edo and Newatia, Karan and Ma, Yiping and Zhong, Ke and Angel, Sebastian and Haeberlen, Andreas},
  booktitle = {Proceedings of the ACM SIGOPS 28th Symposium on Operating Systems Principles},
  pages     = {327--343},
  year      = {2021}
}

@inproceedings{arboretum,
  title     = {Arboretum: A planner for large-scale federated analytics with differential privacy},
  author    = {Margolin, Elizabeth and Newatia, Karan and Luo, Tao and Roth, Edo and Haeberlen, Andreas},
  booktitle = {Proceedings of the 29th Symposium on Operating Systems Principles},
  pages     = {451--465},
  year      = {2023}
}

@article{confidentialVMsExplained,
  author     = {Misono, Masanori and Stavrakakis, Dimitrios and Santos, Nuno and Bhatotia, Pramod},
  title      = {Confidential VMs Explained: An Empirical Analysis of AMD SEV-SNP and Intel TDX},
  year       = {2024},
  issue_date = {December 2024},
  publisher  = {Association for Computing Machinery},
  address    = {New York, NY, USA},
  volume     = {8},
  number     = {3},
  url        = {https://doi.org/10.1145/3700418},
  doi        = {10.1145/3700418},
  journal    = {Proc. ACM Meas. Anal. Comput. Syst.},
  month      = dec,
  articleno  = {36},
  numpages   = {42}
}

@article{sev2020strengthening,
  title   = {Strengthening VM isolation with integrity protection and more},
  author  = {Sev-Snp, AMD},
  journal = {White Paper, January},
  volume  = {53},
  number  = {2020},
  pages   = {1450--1465},
  year    = {2020}
}

@misc{intelTDX,
  author       = {},
  title        = {Intel® Trust Domain Extensions (Intel® TDX)},
  howpublished = {\url{https://www.intel.com/content/www/us/en/developer/tools/trust-domain-extensions/overview.html}},
  year         = 2025,
  note         = {Accessed: 2025-11-17}
}

@misc{armCCA,
  author       = {},
  title        = {Arm Confidential Compute Architecture},
  howpublished = {\url{https://www.arm.com/architecture/ security-features/arm-confidential-compute-architecture}},
  year         = 2025,
  note         = {Accessed: 2025-11-17}
}

@inproceedings{huba2022papaya,
  author    = {Huba, Dzmitry and Nguyen, John and Malik, Kshitiz and Zhu, Ruiyu and Rabbat, Mike and Yousefpour, Ashkan and Wu, Carole-Jean and Zhan, Hongyuan and Ustinov, Pavel and Srinivas, Harish and Wang, Kaikai and Shoumikhin, Anthony and Min, Jesik and Malek, Mani},
  booktitle = {Proceedings of Machine Learning and Systems},
  editor    = {D. Marculescu and Y. Chi and C. Wu},
  pages     = {814--832},
  title     = {PAPAYA: Practical, Private, and Scalable Federated Learning},
  url       = {https://proceedings.mlsys.org/paper_files/paper/2022/file/a8bc4cb14a20f20d1f96188bd61eec87-Paper.pdf},
  volume    = {4},
  year      = {2022}
}

@inproceedings{bonawitz2019googlefl,
  author    = {Bonawitz, Keith and Eichner, Hubert and Grieskamp, Wolfgang and Huba, Dzmitry and Ingerman, Alex and Ivanov, Vladimir and Kiddon, Chlo\'{e} and Kone\v{c}n\'{y}, Jakub and Mazzocchi, Stefano and McMahan, Brendan and Van Overveldt, Timon and Petrou, David and Ramage, Daniel and Roselander, Jason},
  booktitle = {Proceedings of Machine Learning and Systems},
  editor    = {A. Talwalkar and V. Smith and M. Zaharia},
  pages     = {374--388},
  title     = {Towards Federated Learning at Scale: System Design},
  url       = {https://proceedings.mlsys.org/paper_files/paper/2019/file/7b770da633baf74895be22a8807f1a8f-Paper.pdf},
  volume    = {1},
  year      = {2019}
}

@article{paulik2021applefl,
  title   = {Federated Evaluation and Tuning for On-Device Personalization: System Design \& Applications},
  author  = {Matthias Paulik and Matt Seigel and Henry Mason and Dominic Telaar and Joris Kluivers and Rogier van Dalen and Chi Wai Lau and Luke Carlson and Filip Granqvist and Chris Vandevelde and Sudeep Agarwal and Julien Freudiger and Andrew Byde and Abhishek Bhowmick and Gaurav Kapoor and Si Beaumont and Áine Cahill and Dominic Hughes and Omid Javidbakht and Fei Dong and Rehan Rishi and Stanley Hung},
  year    = {2022},
  journal = {arXiv preprint arXiv:2102.08503},
  url     = {https://arxiv.org/pdf/2102.08503.pdf}
}

@techreport{trustinfood2021farmerdata,
  title       = {Farmer Perspectives on Data 2021},
  author      = {Slattery, Drew and Rayburn, Kinsie and Slay, Christy Melhart and Garcia-Moore, Teresa},
  institution = {Trust In Food, a Farm Journal Initiative and The Sustainability Consortium},
  year        = {2021},
  url         = {https://h2020-demeter.eu/wp-content/uploads/2021/06/Farmer-Perspectives-on-Data-2021-1.pdf},
  note        = {Accessed: 2025-12-01}
}

@article{iqbal2025sustainable,
  title     = {Sustainable food systems transformation in the face of climate change: strategies, challenges, and policy implications},
  author    = {Iqbal, Babar and Alabbosh, Khulood Fahad and Jalal, Abdul and Suboktagin, Sultan and Elboughdiri, Noureddine},
  journal   = {Food Science and Biotechnology},
  volume    = {34},
  number    = {4},
  pages     = {871--883},
  year      = {2025},
  publisher = {Springer}
}

@article{dembani2025agricultural,
  title     = {Agricultural data privacy and federated learning: A review of challenges and opportunities},
  author    = {Dembani, Rahool and Karvelas, Ioannis and Akbar, Nur Arifin and Rizou, Stamatia and Tegolo, Domenico and Fountas, Spyros},
  journal   = {Computers and Electronics in Agriculture},
  volume    = {232},
  pages     = {110048},
  year      = {2025},
  publisher = {Elsevier}
}

@techreport{FAO2009,
  author      = {{Food and Agriculture Organization of the United Nations}},
  title       = {How to Feed the World in 2050},
  year        = {2009},
  institution = {FAO},
  url         = {https://www.fao.org/fileadmin/templates/wsfs/docs/expert_paper/How_to_Feed_the_World_in_2050.pdf},
  note        = {Accessed: 2025-12-01}
}

@article{wilgenbusch2022addressing,
  title     = {Addressing new data privacy realities affecting agricultural research and development: A tiered-risk, standards-based approach},
  author    = {Wilgenbusch, James C and Pardey, Philip G and Hospodarsky, Naomi and Lynch, Benjamin J},
  journal   = {Agronomy Journal},
  volume    = {114},
  number    = {5},
  pages     = {2653--2668},
  year      = {2022},
  publisher = {Wiley Online Library}
}

@inproceedings{shokri2017membership,
  title        = {Membership inference attacks against machine learning models},
  author       = {Shokri, Reza and Stronati, Marco and Song, Congzheng and Shmatikov, Vitaly},
  booktitle    = {2017 IEEE symposium on security and privacy (SP)},
  pages        = {3--18},
  year         = {2017},
  organization = {IEEE}
}

@article{so2021turbo,
  title     = {Turbo-aggregate: Breaking the quadratic aggregation barrier in secure federated learning},
  author    = {So, Jinhyun and G{\"u}ler, Ba{\c{s}}ak and Avestimehr, A Salman},
  journal   = {IEEE Journal on Selected Areas in Information Theory},
  volume    = {2},
  number    = {1},
  pages     = {479--489},
  year      = {2021},
  publisher = {IEEE}
}

@misc{katacontainers,
  title        = {{Kata Containers}},
  howpublished = {\url{https://katacontainers.io/}},
  note         = {Accessed: 2025-12-06},
  year         = {2017},
  author       = {{Kata Containers Project}},
  organization = {Open Infrastructure Foundation}
}

@misc{confidentialcontainers,
  author       = {{Confidential Containers Project}},
  title        = {{Confidential Containers}},
  howpublished = {\url{https://confidentialcontainers.org/}},
  note         = {Accessed: 2025-12-06},
  year         = {2022},
  organization = {Confidential Computing Consortium}
}

@misc{nguyen2022federatedlearningbufferedasynchronous,
  title         = {Federated Learning with Buffered Asynchronous Aggregation},
  author        = {John Nguyen and Kshitiz Malik and Hongyuan Zhan and Ashkan Yousefpour and Michael Rabbat and Mani Malek and Dzmitry Huba},
  year          = {2022},
  eprint        = {2106.06639},
  archiveprefix = {arXiv},
  primaryclass  = {cs.LG},
  url           = {https://arxiv.org/abs/2106.06639}
}

@inproceedings{chen2020asynchronous,
  title        = {Asynchronous online federated learning for edge devices with non-iid data},
  author       = {Chen, Yujing and Ning, Yue and Slawski, Martin and Rangwala, Huzefa},
  booktitle    = {2020 IEEE International Conference on Big Data (Big Data)},
  pages        = {15--24},
  year         = {2020},
  organization = {IEEE}
}

@inproceedings{fedavg,
  title        = {Communication-efficient learning of deep networks from decentralized data},
  author       = {McMahan, Brendan and Moore, Eider and Ramage, Daniel and Hampson, Seth and y Arcas, Blaise Aguera},
  booktitle    = {Artificial intelligence and statistics},
  pages        = {1273--1282},
  year         = {2017},
  organization = {PMLR}
}

@misc{farmbot,
  title        = {Farmbot | Open-source CNC Farming },
  howpublished = {\url{https://farm.bot/}},
  author       = {FarmBot Inc.},
  note         = {Accessed: 2025-12-06}
}

@misc{cimis_stations2025,
  title        = {Stations - California Irrigation Management Information System (CIMIS)},
  howpublished = {\url{https://cimis.water.ca.gov/Stations.aspx}},
  note         = {Accessed: 2025-12-11},
  year         = {2025},
  organization = {California Department of Water Resources, State of California},
  url          = {https://cimis.water.ca.gov/Stations.aspx}
}

@inproceedings{dp,
  title        = {Calibrating noise to sensitivity in private data analysis},
  author       = {Dwork, Cynthia and McSherry, Frank and Nissim, Kobbi and Smith, Adam},
  booktitle    = {Theory of cryptography conference},
  pages        = {265--284},
  year         = {2006},
  organization = {Springer}
}

@book{r_squre,
  title     = {Applied regression analysis},
  author    = {Draper, Norman R and Smith, Harry},
  volume    = {303},
  year      = {1998},
  publisher = {John Wiley \& Sons}
}

@misc{unfss2_highlights,
  author       = {{UNFCCC}},
  title        = {{United Nations Food Systems Summit (+2) Highlights}},
  howpublished = {\url{https://unfccc.int/sites/default/files/resource/unfss2_highlights.pdf}},
  note         = {Briefing Note, accessed: December 6, 2025},
  year         = {2023}
}

@misc{APLUlandgrant,
  author       = {The Association of Public and Land-grant Universities},
  title        = {What is a land-grant university?},
  howpublished = {\url{https://www.aplu.org/about-us/history-of-aplu/what-is-a-land-grant-university/}},
  note         = {FAQ, accessed: May 28, 2026}
}

@article{ponomareva2023howtodpfy,
  title   = {How to {DP}-fy {ML}: A Practical Guide to Machine Learning with Differential Privacy},
  author  = {Ponomareva, Natalia and Hazimeh, Hussein and Kurakin, Alex and Xu, Zheng and Denison, Carson and McMahan, H. Brendan and Vassilvitskii, Sergei and Chien, Steve and Thakurta, Abhradeep Guha},
  journal = {Journal of Artificial Intelligence Research},
  volume  = {77},
  pages   = {1113--1201},
  year    = {2023}
}

@article{tilman2011global,
  title     = {Global food demand and the sustainable intensification of agriculture},
  author    = {Tilman, David and Balzer, Christian and Hill, Jason and Befort, Belinda L.},
  journal   = {Proceedings of the National Academy of Sciences},
  volume    = {108},
  number    = {50},
  pages     = {20260--20264},
  year      = {2011},
  month     = {December},
  publisher = {National Academy of Sciences}
}

@inproceedings{vasisht2017farmbeats,
  title     = {{FarmBeats}: An {IoT} Platform for Data-Driven Agriculture},
  author    = {Vasisht, Deepak and Kapetanovic, Zerina and Won, Jongho and Jin, Xinxin and Chandra, Ranveer and Sinha, Sudipta and Kapoor, Ashish and Sudarshan, Madhusudhan and Stratman, Sean},
  booktitle = {14th USENIX Symposium on Networked Systems Design and Implementation (NSDI 17)},
  pages     = {515--529},
  address   = {Boston, MA},
  month     = {March},
  year      = {2017},
  publisher = {USENIX Association}
}

@inproceedings{steininger2023cropandweed,
  title     = {The {CropAndWeed} Dataset: A Multi-Modal Learning Approach for Semantic Segmentation},
  author    = {Steininger, Daniel and Trondl, Andreas and Croonenborghs, Gerardus and Simon, Julia and Widhalm, Verena},
  booktitle = {Proceedings of the IEEE/CVF Winter Conference on Applications of Computer Vision (WACV)},
  pages     = {3729--3738},
  year      = {2023}
}

@article{hughes2015plantvillage,
  title   = {An open access repository of images on plant health to enable the development of mobile disease diagnostics},
  author  = {Hughes, David P. and Salath{\'e}, Marcel},
  journal = {arXiv preprint arXiv:1511.08060},
  year    = {2015}
}

@article{zou2004statistical,
  title   = {Statistical validation of image segmentation quality based on a spatial overlap index},
  author  = {Zou, Kelly H. and Warfield, Simon K. and Bharatha, Aditya and Tempany, Clare M. C. and Kaus, Michael R. and Haker, Steven J. and Wells, William M. and Jolesz, Ferenc A. and Kikinis, Ron},
  journal = {Academic Radiology},
  volume  = {11},
  number  = {2},
  pages   = {178--189},
  year    = {2004},
  month   = feb,
  doi     = {10.1016/S1076-6332(03)00671-8},
  pmid    = {14974593},
  pmcid   = {PMC1415224}
}

@inproceedings{Rubambiza2022seamless,
  author    = {Rubambiza, Gloire and Sengers, Phoebe and Weatherspoon, Hakim},
  title     = {Seamless Visions, Seamful Realities: Anticipating Rural Infrastructural Fragility in Early Design of Digital Agriculture},
  year      = {2022},
  isbn      = {9781450391573},
  publisher = {Association for Computing Machinery},
  address   = {New York, NY, USA},
  url       = {https://doi.org/10.1145/3491102.3517579},
  doi       = {10.1145/3491102.3517579},
  booktitle = {Proceedings of the 2022 CHI Conference on Human Factors in Computing Systems},
  articleno = {451},
  numpages  = {15},
  location  = {New Orleans, LA, USA},
  series    = {CHI '22}
}

@inbook{Cremonesi2025,
  author    = {Cremonesi, Francesco
               and Vesin, Marc
               and Cansiz, Sergen
               and Bouillard, Yannick
               and Balelli, Irene
               and Innocenti, Lucia
               and Taiello, Riccardo
               and Silva, Santiago
               and Ayed, Samy-Safwan
               and {\"O}nen, Melek
               and Orlhac, Fanny
               and Nioche, Christophe
               and Houis, Bastien
               and Modzelewski, Romain
               and Lapel, Nathan
               and Schiappa, Renaud
               and Humbert, Olivier
               and Lorenzi, Marco},
  editor    = {Rehman, Muhammad Habib ur
               and Gaber, Mohamed Medhat},
  title     = {Fed-BioMed: Open, Transparent and Trusted Federated Learning for Real-world Healthcare Applications},
  booktitle = {Federated Learning Systems: Towards Privacy-Preserving Distributed AI},
  year      = {2025},
  publisher = {Springer Nature Switzerland},
  address   = {Cham},
  pages     = {19--41},
  isbn      = {978-3-031-78841-3},
  doi       = {10.1007/978-3-031-78841-3_2},
  url       = {https://doi.org/10.1007/978-3-031-78841-3_2}
}

@misc{k3s,
  author = {{K3s Project Authors}},
  title  = {K3s - Lightweight Kubernetes},
  url    = {https://docs.k3s.io/},
  note   = {Accessed: 2026-06-08}
}

@article{10.3389/fsufs.2022.903230,
  author  = {Kaur, Jasmin  and Hazrati Fard, Seyed Mehdi  and Amiri-Zarandi, Mohammad  and Dara, Rozita },
  title   = {Protecting farmers' data privacy and confidentiality: Recommendations and considerations},
  journal = {Frontiers in Sustainable Food Systems},
  volume  = {Volume 6 - 2022},
  year    = {2022},
  url     = {https://www.frontiersin.org/journals/sustainable-food-systems/articles/10.3389/fsufs.2022.903230},
  doi     = {10.3389/fsufs.2022.903230},
  issn    = {2571-581X}
}

@inproceedings{giersfeld2026breakfast,
  title     = {{BreakFAST: Confused Deputy Attack on Infinity Fabric to Break AMD SEV-SNP}},
  author    = {Philipp Giersfeld and Benedict Schl\"uter and Shweta Shinde},
  booktitle = {47th IEEE Symposium on Security and Privacy (S\&P 26)},
  year      = {2026},
  month     = may,
  address   = {San Francisco, CA},
  publisher = {IEEE}
}

@inproceedings{schlueter2026fabricked,
  title     = {{Fabricked: Misconfiguring Infinity Fabric to Break AMD SEV-SNP}},
  author    = {Benedict Schlüter and Christoph Wech and Shweta Shinde},
  booktitle = {35th USENIX Security Symposium (USENIX Security 26)},
  year      = {2026},
  month     = aug,
  address   = {Baltimore, MD},
  publisher = {USENIX Association}
}

@inproceedings{schlueter2026staleus,
  title     = {{Staleus: Breaking AMD SEV-SNP via Memory Incoherence}},
  author    = {Benedict Schlüter and Shweta Shinde},
  booktitle = {35th USENIX Security Symposium (USENIX Security 26)},
  year      = {2026},
  month     = aug,
  address   = {Baltimore, MD},
  publisher = {USENIX Association}
}

@inproceedings{kairouz2021practical,
  author    = {Kairouz, Peter and McMahan, H. Brendan and Song, Shuang and Thakkar, Om and Thakurta, Abhradeep and Xu, Zheng},
  title     = {{Practical and Private (Deep) Learning without Sampling or Shuffling}},
  booktitle = {Proceedings of the 38th International Conference on Machine Learning (ICML)},
  year      = {2021},
  pages     = {5213--5225}
}

@inproceedings{mironov2017renyi,
  author    = {Mironov, Ilya},
  title     = {{R{\'e}nyi Differential Privacy}},
  booktitle = {2017 IEEE 30th Computer Security Foundations Symposium (CSF)},
  year      = {2017},
  pages     = {263--275},
  publisher = {IEEE},
  doi       = {10.1109/CSF.2017.11}
}

@misc{flower,
  title  = {{Flower: A Friendly Federated Learning Framework}},
  author = {{Flower Labs}},
  year   = {2024},
  url    = {https://flower.ai/},
  note   = {Accessed: 2026-06-09}
}

@misc{fedml,
  title  = {{FedML}},
  author = {{FedML, Inc.}},
  year   = {2024},
  url    = {https://github.com/FedML-AI/FedML},
  note   = {Accessed: 2026-06-09}
}

@misc{flower-dp,
  title  = {{Flower Framework Documentation: Use Differential Privacy}},
  author = {{Flower Labs}},
  year   = {2024},
  url    = {https://flower.ai/docs/framework/how-to-use-differential-privacy.html},
  note   = {Accessed: 2026-06-09}
}

@misc{nvflare-cc,
  title  = {{NVIDIA FLARE: Confidential Federated AI Documentation}},
  author = {{NVIDIA Corporation}},
  year   = {2024},
  url    = {https://nvflare.readthedocs.io/en/main/user_guide/confidential_computing/index.html},
  note   = {Accessed: 2026-06-09}
}

@book{wolf1997privatization,
  title={Privatization of information and agricultural industrialization},
  author={Wolf, Steven A},
  year={1997},
  publisher={CRC Press}
}
